\documentclass[12pt]{article}
\pdfoutput=1

\usepackage{amsmath,amssymb,amscd}
\usepackage{listings}
\usepackage{caption}
\usepackage{dsfont}
\usepackage{slashed}
\usepackage{color}

\usepackage[pdftex]{graphicx}
\usepackage{epstopdf}
\usepackage{subfigure}
\usepackage{epsfig}
\usepackage{listings}
\usepackage{caption}
\usepackage{cite}

\usepackage{multirow}

\newcommand{\bea}{\begin{align}}
\newcommand{\eea}{\end{align}}
\newcommand{\beq}{\begin{equation}}
\newcommand{\eeq}{\end{equation}}
\newcommand{\nbea}{\begin{align*}}
\newcommand{\neea}{\end{align*}}
\newcommand{\nbeq}{\begin{equation*}}
\newcommand{\neeq}{\end{equation*}}
\newcommand{\bear}{\begin{eqnarray}}  
\newcommand{\eear}{\end{eqnarray}}  

 \newcommand{\twomatrix}[1]{\left(\begin{array}{cc} #1 \end{array}\right) }

 \newcommand{\identity}{\mathds{1}}
 
 \newcommand{\tb}{\tan \beta}

\numberwithin{equation}{section}
\begin{document}

%\begin{tiFCC-eeage}

%\pagestyle{empty}

\baselineskip=21pt
\rightline{KCL-PH-TH/2015-17, LCTS/2015-08, CERN-PH-TH/2015-072}
\vskip 1in

\begin{center}

{\large {\bf Comparing EFT and Exact One-Loop Analyses of Non-Degenerate Stops}}
%{\large {\bf One-Loop Effective Action Analysis of Non-Degenerate Stops }}
%{\large {\bf One-Loop Effective Action Analysis and Constraints on Light Stops }}
%{\large {\bf EFT Constraints on Non-Degenerate Stops in the Universal One-Loop Effective Action }}

\vskip 0.6in

{\bf Aleksandra~Drozd}$^{1}$,
 {\bf John~Ellis}$^{1,2}$,
 {\bf J\'er\'emie~Quevillon}$^{1}$
and {\bf Tevong~You}$^{1}$

\vskip 0.4in

{\small {\it

$^1${Theoretical Particle Physics and Cosmology Group, Physics Department, \\
King's College London, London WC2R 2LS, UK}\\
$^2${TH Division, Physics Department, CERN, CH-1211 Geneva 23, Switzerland}
}}

\vskip 0.75in

{\bf Abstract}

\end{center}

\baselineskip=18pt \noindent

%%%%%%%%%%%%%%%%%%%%%%%%%%%%%%%%%%%%%%%%%%%%%%%%%

{\small

We develop a universal approach to the one-loop effective field theory (EFT) 
using the Covariant Derivative Expansion (CDE) method.
We generalise previous results to include broader classes of UV models,
showing how expressions previously obtained
assuming degenerate heavy-particle masses can be extended to non-degenerate cases. We apply our method
to the general MSSM with non-degenerate stop squarks, illustrating our approach with
calculations of the coefficients of dimension-6 operators contributing to the $hgg$ and $h\gamma\gamma$ couplings,
and comparing with exact calculations of one-loop Feynman diagrams. 
We then use present and projected future sensitivities to these operator coefficients
to obtain present and possible future indirect constraints on stop masses.
The current sensitivity is already comparable to that of direct LHC searches, and future FCC-ee measurements 
could be sensitive to stop masses above a TeV. The universality of our one-loop EFT approach
facilitates extending these constraints to a broader class of UV models.

%%%%%%%%%%%%%%%%%%%%%%%%%%%%%%%%%%%%%%%%%%%%%%%%

\vskip 1in

\leftline{April 2015}

%\end{tiFCC-eeage}
\newpage

%\tableofcontents

%*******************************************************************************************
\section{Introduction}

In view of the overall consistency between the current measurements of
particle properties and predictions in the Standard Model (SM), a common
approach to the analysis of present and prospective future data is to
describe them via an effective field theory (EFT) in which the
renormalizable SM $d=4$ Lagrangian is supplemented with higher-dimensional
terms composed from SM fields~\cite{buchmullerwyler, eomreduction}. To the
extent that this new physics has a mass scale that is substantially higher
than the energy scale of the available measurements~\cite{decoupling},
the EFT approach is a powerful way to constrain
possible new physics beyond the SM (BSM) that is model-independent~\cite{eftconstraints, ESYhiggs, ESY}.
The $d=6$ operators in this Effective SM (ESM) were first classified in~\cite{buchmullerwyler}\footnote{This EFT approach
that we follow, in which the $SU(2)_L \times U(1)_Y$ electroweak symmetry is linearly realized,
is to be distinguished from a non-linear EFT based on the chiral electroweak Lagrangian~\cite{nonlinearEFT}
and the more general anomalous coupling framework of a $U(1)_\text{EM}$ effective Lagrangian~\cite{anomalousEFT}.}, 
with a complete basis using equations of motion to eliminate redundancies~\cite{eomreduction} 
being first presented in~\cite{GIMR}. There have been many studies of various aspects of these 
dimension-6 operators~\footnote{See~\cite{buchmullerwyler, eomreduction, earlyeft} for some 
examples of earlier work and~\cite{GIMR, eftconstraints, ESYhiggs, ESY, recenteftpapers, HLM1,HLM2, craigetal, extendedHiggsEFT} 
for a sampling of more recent studies.}, and a short review can be found in~\cite{EFTreview}. 

The EFT approach may well be a good approximation if the new physics affects 
precision observables at the tree level, or if it is strongly-interacting. In these
cases the new physics mass scale is likely to be relatively high, and considering
the lowest-dimensional EFT operators may well be sufficient. However, the EFT approach may 
have limitations if the new physics has effects only at the loop level, or is weakly
interacting. In these cases, the EFT approach may be sensitive only to new physics at
some relatively low mass scale, and the new physics effects may not be characterised well
by considering simply the lowest-dimensional EFT operators.

Examples in the first, `safer' category may include certain models with extended
Higgs sectors~\cite{extendedHiggsEFT}, such as two-Higgs-doublet models, or some composite models.
Examples in the second category may include the loop effects of supersymmetric
models. However, even in this case it is possible that precision electroweak and
Higgs data may provide interesting constraints on the possible masses of stop squarks,
which have relatively large Yukawa couplings to the SM Higgs field. In particular, the EFT
approach may be useful in the framework of `natural' supersymmetric models with stops
that have masses above 100~GeV but still relatively light compared to other
supersymmetric particles.

Important steps towards the calculation of loop effects and the simplification of their
matching with EFT coefficients have been taken recently by Henning, Lu and Murayama (HLM)~\cite{HLM1,HLM2}.
In particular, they use a covariant-derivative expansion (CDE)~\cite{Gaillard,Cheyette} to characterise new-physics
effects via the evaluation of the one-loop effective action. They apply these techniques to
derive universal results and also study some explicit models including electroweak triplet
scalars, an extra electroweak scalar doublet, and light stops within the minimal
supersymmetric extension of the SM (MSSM), as well as some other models. They
also discuss electroweak precision observables, triple-gauge couplings and Higgs decay
widths and production cross sections~\cite{HLM2}, and have used their results to derive indicative
constraints on the basis of present and future data~\cite{HLM1}.

In this paper we discuss aspects of the applicability of the EFT approach
to models with relatively light stops, exploring in more depth some issues arising
from the work of HLM~\cite{HLM1,HLM2}. As they discuss, using the CDE and
the one-loop effective action is more elegant and less time-consuming than a
complete one-loop Feynman diagram computation. On the other hand, they
applied their approach to models with degenerate soft supersymmetry-breaking
terms for the stop squarks, and we show how to extend their approach to the
non-degenerate case, with specific applications to the dimension-6 operators
that contribute to the $hgg$ and $h\gamma\gamma$ couplings. Our extension of the CDE approach would also permit
applications to a wider class of ultra-violet (UV) extensions of the SM and
other EFT operators.

Another important aspect of our work is a comparison of the EFT results with the
corresponding full one-loop Feynman diagram calculations also in the 
non-degenerate case, so as to assess the accuracy of
the EFT approach for analysing present and future data. 

In a recent paper,
together with Sanz, two of us (JE and TY) made a global fit to dimension-6
EFT operator coefficients including electroweak precision data, LHC
measurements of triple-gauge couplings, Higgs rates and production
kinematics~\cite{ESY}. Here we use this global fit to constrain the stop mass 
$m_{\tilde t}$ and the mixing parameter $X_t$,
comparing results obtained using the EFT with those using the full one-loop
diagrammatic calculation. The bounds on $m_{\tilde t}$ and $X_t$ are strongly
correlated, and we find that the EFT approach may yield 
quite accurate constraints for the limits of larger $m_{\tilde t}$ and $X_t$.
However, there are substantial differences from the full diagrammatic result
for smaller $m_{\tilde t}$ and $X_t$. In this case the diagrammatic approach
gives indirect constraints on the stop squark that are quite competitive with
direct experimental searches at the LHC. We also explore the possible accuracy
of the EFT for possible future data sets, including those obtainable from the LHC
and possible $e^+ e^-$ colliders~\footnote{For previous analyses, see~\cite{HLM1,FM, FMW}.}.
For example, possible FCC-ee measurements~\cite{FCC-ee}
may be sensitive indirectly to stop masses $\gtrsim 1$~TeV.

The layout of this paper is as follows. In Section~2 we introduce the
covariant derivative expansion (CDE) and discuss its application to the
one-loop effective action, highlighting how the HLM approach~\cite{HLM1,HLM2} may be
extended to the case of non-degenerate squarks. As we discuss, one
way to achieve this is to use the Baker-Campbell-Hausdorff (BCH) theorem to rearrange the
one-loop effective action, and another is to introduce an auxiliary expansion variable.
Results obtained by these two methods agree, and
are also consistent with the full one-loop Feynman diagram result presented
in Section~3. Analyses of the current data in the frameworks of the EFT
and the diagrammatic approach are presented in Section~4, and their
results compared. Studies of the possible sensitivities of future
measurements at the ILC and FCC-ee are presented in Section~5, and Section~6 discusses
our conclusions and possible directions for future work.

\section{The Covariant Derivative Expansion and the One-Loop Effective Action}

The one-loop effective action may be obtained by integrating out directly the heavy particles in the path integral 
using the saddle-point approximation of the functional integral. The contributions to operators involving only light fields 
can be evaluated by various expansion methods for the application of the path integral. Here we follow the Covariant Derivative Expansion (CDE), 
a manifestly gauge-invariant method first introduced in the 1980s by
Gaillard~\cite{Gaillard} and Cheyette~\cite{Cheyette}, 
and recently applied to the Effective SM (ESM) by Henning, Lu and Murayama (HLM)~\cite{HLM2}~\footnote{We thank Herm\`{e}s B\'{e}lusca-Ma\"{i}to for pointing out to us another recent paper that computes the one-loop effective action for certain dimension-6 QCD operators~\cite{tokyodim6}.}. The latter
provide, in particular, universal results for operators up to dimension-6 in the form of a one-loop effective Lagrangian
with coefficients evaluated via momentum integrals. This approach applies generally, 
and greatly simplifies the matching to UV models, since it avoids the necessity of recalculating 
one-loop Feynman diagrams for every model. However, HLM assume a degenerate mass matrix,
which may not be the case in general, as for example in the `natural' MSSM with light stops. 
We show here how their results may be extended to the non-degenerate case for the 
one-loop effective Lagrangian terms involved in the dimension-6 operators affecting the
$hgg$ and $h\gamma\gamma$ couplings, with application to the case of non-degenerate stops and sbottoms.

\subsection{The Non-Degenerate One-Loop Effective Lagrangian}

We consider a generic Lagrangian consisting of the SM part with complex heavy scalar fields arranged in a multiplet $\Phi$, 
\begin{equation}
\mathcal{L}_\text{UV} = \mathcal{L}_\text{SM} + (\Phi^\dagger F(x) + \text{h.c.}) + \Phi^\dagger(P^2 - M^2 - U(x))\Phi + \mathcal{O}(\Phi^3) \, ,
\label{eq:lagrangianUV}
\end{equation}
where $P \equiv iD_\mu$, with $D_\mu$ the gauge-covariant derivative, 
$F(x)$ and $U(x)$ are combinations of SM fields coupling linearly and quadratically respectively to $\Phi$, 
and $M$ is a diagonal mass matrix. The path integral over $\Phi$ may be computed by expanding the action 
around the minimum with respect to $\Phi$, so that the linear terms give the tree-level effective Lagrangian
upon substituting the equation of motion for $\Phi$:
\begin{equation*}
\mathcal{L}^\text{eff}_\text{tree} = \sum_{n=0} F^\dagger M^{-2} [(P^2-U)M^{-2}]^n F + \mathcal{O}(\Phi^3) \, ,
\end{equation*}
whereas the quadratic terms are responsible for the one-loop part of the effective Lagrangian. 
After evaluating the functional integral and Fourier transforming to momentum space, this can be written in the form
\begin{equation*}
\mathcal{L}^\text{eff}_\text{1-loop} = i \int \frac{d^4q}{(2\pi)^4} \text{Tr} \ln(-(P_\mu-q_\mu)^2 + M^2 + U) \, .
\end{equation*}
It is convenient, before expanding the logarithm, to shift the momentum
using the covariant derivative, by inserting factors of $e^{\pm P_\mu\partial/\partial q_\mu}$: 
\begin{equation*}
\mathcal{L}^\text{eff}_\text{1-loop} = i \int \frac{d^4q}{(2\pi)^4} \text{Tr} \ln[e^{P_\mu\partial/\partial q_\mu}(-(P_\mu-q_\mu)^2 + M^2 + U) e^{-P_\mu\partial/\partial q_\mu} ] \, .
\end{equation*}
This choice ensures a convergent expansion while the calculation of operators remains manifestly 
gauge-invariant throughout~\footnote{We refer the reader to~\cite{Gaillard, Cheyette, HLM2} for technical details
and discussions of the CDE method.}. The result is a series involving gauge field strengths, 
covariant derivatives and SM fields encoded in the matrix $U(x)$:
\begin{equation*}
\mathcal{L}^\text{eff}_\text{1-loop} = i  \int \frac{d^4q}{(2\pi)^4} \text{Tr} \ln[-(\tilde{G}_{\nu\mu}\partial/\partial q_\mu + q_\mu)^2 + M^2 + \tilde{U}]	\, ,
\end{equation*}
where
\begin{align*}
\tilde{G}_{\nu\mu} &\equiv \sum_{n=0} \frac{n+1}{(n+2)!}[P_{\alpha_1},[...[P_{\alpha_n},G^\prime_{\nu\mu}]]]\frac{\partial^n}{\partial q_{\alpha_1} ... q_{\alpha_n}} 	\, , \\
\tilde{U} &=\sum_{n=0} \frac{1}{n!}[P_{\alpha_1},[...[P_{\alpha_n},U]]]	\, .
\end{align*}
Here we defined $G^\prime_{\nu\mu} \equiv -iG_{\nu\mu}$ with the field strength given by 
$[P_\nu,P_\mu] = -G^\prime_{\nu\mu}$. 
It is convenient to group together the terms involving momentum derivatives:
\begin{equation*}
\mathcal{L}^\text{eff}_\text{1-loop} = i  \int \frac{d^4q}{(2\pi)^4} \text{Tr} \ln(A + B)	\, ,
\end{equation*}
where
\begin{align}
A &\equiv -\{q_\mu, \tilde{G}_{\nu\mu}\}\frac{\partial}{\partial q_\nu} - \tilde{G}_{\nu\mu}\tilde{G}_{\alpha\mu}\frac{\partial^2}{\partial q_\nu q_\alpha} + \delta\tilde{U}	
\label{eq:AandB}
\, , \\
B &\equiv -q^2 + M^2 + U \nonumber \, ,
\end{align}
and we have separated $\tilde{U} = U + \delta\tilde{U}$. 

Expanding the logarithm using the 
Baker-Campbell-Hausdorff (BCH) formula gives
\begin{equation*}
\ln(A+B) = \ln(B) + \ln(1+B^{-1}A) + \frac{1}{2}[\ln B, \ln(1+B^{-1}A)] + \frac{1}{12}[\ln B, [\ln B, \ln(1+B^{-1}A)]] + ...	
\end{equation*}
and, using the identity $[\ln X, Y] = \sum_{n=1} \frac{1}{n} X^{-n} L^n_X Y$, where $L_X Y \equiv [X,Y]$, we see that all possible gauge-invariant operators are obtained by evaluating commutators of $A$ and $B$. 

As an example, we compute the term contributing to the dimension-6 operator affecting Higgs production by gluon fusion:
\begin{equation*}
\mathcal{O}_g = g_3^2 |H^2| G^a_{\mu\nu}{G^a}^{\mu\nu} \, .
\end{equation*}
The calculation can be organised by writing $A$ as a series in momentum derivatives, 
\begin{equation*}
A = \sum_{n=1} A^{\alpha_1 ... \alpha_n}_n \frac{\partial^n}{\partial q_{\alpha_1} ... \partial q_{\alpha_n}}	 
= A_1^{\alpha_1}\frac{\partial}{\partial q_{\alpha_1}} + A_2^{\alpha_1\alpha_2}\frac{\partial^2}{\partial q_{\alpha_1 q_{\alpha_2}}} + \text{...} \, ,
\end{equation*}
where each term is obtained by substituting $\tilde{G}$ and $\tilde{U}$ in Eq.~\ref{eq:AandB}. Here we require only the part $A_2^{\alpha_1\alpha_2} \supset -\frac{1}{4}G^\prime_{\alpha_1\mu}G^\prime_{\alpha_2\mu}$, together with the following commutators:
\begin{equation*}
i  \int \frac{d^4q}{(2\pi)^4} \text{Tr} \ln(A+B) \supset i \int \frac{d^4q}{(2\pi)^4} \text{Tr}\left( \frac{1}{2} B^{-2}[B,A] + \frac{1}{3}B^{-3}[B,[B,A]] \right)	\, .
\end{equation*} 
We note that $M$ and $U$ are $n \times n$ matrices that do not commute in general,
which motivates the use of the BCH expansion, first applied to the CDE in~\cite{Cheyette}.
Evaluating the commutators we find
\begin{equation*}
\mathcal{L}^\text{eff}_\text{1-loop} \supset i \int \frac{d^4q}{(2\pi)^4} \text{Tr}\left\{ B^{-2}\left(-\frac{1}{4}G^\prime_{\nu\mu}{G^\prime}^{\nu\mu} \right) -\frac{8}{3}q_\alpha q_\nu B^{-3}\left(-\frac{1}{4}{G^\prime}^\alpha {}_\mu{G^\prime}^{\nu\mu} \right) \right\}
\end{equation*}
and using $B^{-1} = -\Delta\sum_{n=0}(\Delta U)^n$, where $\Delta \equiv 1/(q^2 - M^2)$, 
we see that to obtain operators up to dimension 6 requires retaining up to two powers of $U$,
so that we have traces of the form
\begin{align*}
\text{Tr}(\Delta^a U {G^\prime}^\alpha{}_\mu{G^\prime}^{\nu\mu}) &= \sum_{i=1}^n \left( \Delta_i^a U_{ii} {G^\prime_i}^\alpha {}_\mu{G^\prime_i}^{\nu\mu} \right) \, , \\
\text{Tr}(\Delta^a U \Delta^b U \Delta^c {G^\prime}^\alpha{}_\mu{G^\prime}^{\nu\mu}) &= \sum_{i=1}^n \sum_{j=1}^n \left(\Delta^{a+c}_i \Delta^b_{j}  U_{ij} U_{ji}{G^\prime_i}^\alpha {}_\mu{G^\prime_i}^{\nu\mu} \right)	 \, .
\end{align*}
Here we assume $G^\prime = \text{diag}(G^\prime_1, ... , G^\prime_n)$
and $\Delta = \text{diag}(\Delta_1, ... , \Delta_n)$, where $\Delta_i \equiv 1/(q^2 - m_i^2)$, 
and $U$ is a general $n \times n $ matrix. To evaluate the momentum integrals of arbitrary powers of mixed propagators we
need to combine them using Feynman parameters:
\begin{equation*}
\int \frac{d^4q}{(2\pi)^4} q^l \Delta^a_i \Delta^b_j = \frac{(a+b+1)!}{(a-1)!(b-1)!} \int_0^1 dz_i dz_j \left[ z_i^{a-1} z_j^{b-1} \left(\int \frac{d^4q}{(2\pi)^4} q^l \Delta^{a+b}_{ij} \right) \delta(1-z_i-z_j) \right] \, ,
\end{equation*}
where $\Delta_{ij} \equiv 1 / (q^2 - m_i^2 z_i -  m_j^2  z_j)$. Taking care in applying the $\delta-$function 
in the summation over the matrix indices, we finally obtain the following expression valid in the case of a 
non-degenerate mass matrix:
\begin{equation}
\mathcal{L}^\text{eff}_\text{1-loop} \supset \frac{1}{(4\pi)^2}\left[ -\frac{1}{12} \sum_{i=1}^n \left( \frac{U_{ii}}{m_i^2} {G^\prime_i}_{\mu\nu}{G^\prime_i}^{\mu\nu} \right) + \frac{1}{24}\sum_{i=1}^n \sum_{j=1}^n \left( \frac{U_{ij}U_{ji}}{m_i^2m_j^2}{G^\prime_i}_{\mu\nu}{G^\prime_i}^{\mu\nu} \right) \right ]	\, .
\label{eq:universalGGdiag}
\end{equation}

We have checked this result by extending the log-expansion method of~\cite{HLM2} to the non-degenerate case by
introducing an auxiliary parameter $\xi$ and then differentiating under the integral sign:
\begin{align*}
\mathcal{L}^\text{eff}_\text{1-loop} &= i  \int \frac{d^4q}{(2\pi)^4} \text{Tr} \ln[-(\tilde{G}_{\nu\mu}\partial/\partial q_\mu + q_\mu)^2 + \xi M^2 + \tilde{U}] \\
&=  i  \int \frac{d^4q}{(2\pi)^4} \int d\xi \text{Tr}\left(\frac{1}{A+U - \Delta^{-1}_\xi}M^2 \right)	\, ,
\end{align*}
where $\Delta^\xi \equiv 1 / (q^2 - \xi M^2)$ and $\xi$ is set to 1 at the end of the calculation. The expansion then reads
\begin{align*}
\mathcal{L}^\text{eff}_\text{1-loop} = i  \int \frac{d^4q}{(2\pi)^4} \int d\xi \text{Tr}\left\{ \sum_{n=0}^\infty \left[-\Delta^\xi(A+U)\right]^n\Delta^\xi M^2 \right\}	\, ,
\end{align*}
and yields the same result as in (\ref{eq:universalGGdiag}), demonstrating the consistency of our approach.

In general the field strength matrix $G_{\mu\nu}$ may not be diagonal, as for example when the 
$\Phi$ multiplet contains an $SU(2)_L$ doublet and singlet, so that we have a $2\times 2$ non-diagonal 
sub-matrix $W_{\mu\nu}^a\tau^a$ involving the weak gauge bosons $W^a_{\mu}$. 
The relevant non-degenerate one-loop effective Lagrangian terms then generalise to the universal expression
that can be found in~\cite{DEQY2}~\footnote{The previous version of this paper contained an erroneous expression that was correct under the assumption of $G^\prime_{\mu\nu}$ commuting with $M$ and $U$, but not for the fully general case. However this does not affect any of our results.}.

\subsection{A Light Stop in the $hgg$ and $h\gamma\gamma$ Couplings}

The result of the CDE expansion is universal in the sense that all the
UV information is encapsulated in the $U, M$ matrices and the $P_\mu$ covariant derivative, 
while the operator coefficients are determined by integrals over momenta that are performed once and for all. 
The simplicity of this approach is illustrated by integrating out stops in the MSSM,
whose leading-order contribution necessarily appears at one-loop due to R-parity. Since gluon fusion in the 
SM also occurs at one-loop and currently provides the strongest constraint on any dimension-6 operator in the Higgs sector, 
we first calculate its Wilson coefficient within the EFT framework. Later we extend the calculation to
the the dimension-6 operators contributing to the $h\gamma\gamma$ coupling, and comment on the extension to other dimension-6 operators.

The $M$ and $U$ matrices are given by the quadratic stop term in the MSSM Lagrangian,
\begin{equation*}
\mathcal{L}_\text{MSSM} \supset \Phi^\dagger (M^2 + U(x)) \Phi	\, ,
\end{equation*}
where $ \Phi = (\tilde{Q} \,, \tilde{t}_R^*)$, and 
\begin{equation*}
M^2 = \twomatrix{ m^2_{\tilde{Q}} & 0 \\ 0 & m^2_{\tilde{t}_R} } \, , 
\end{equation*}
\begin{equation*}
U =\twomatrix{
(h_t^2 + \frac{1}{2}g_2^2c^2_\beta)\tilde{H}\tilde{H}^\dagger+\frac{1}{2}g_2^2 s^2_\beta HH^\dagger - \frac{1}{2}(g_1^2 Y_{\tilde{Q}} c_{2\beta} + \frac{1}{2} g_2^2)|H|^2 
&
h_t X_t \tilde{H} 
\\ 
h_t X_t \tilde{H}^\dagger
&
(h_t^2 - \frac{1}{2}g_1^2 Y_{\tilde{t}_R} c_{2\beta})|H|^2
} 	\, .
\end{equation*}
Here we have defined $\tilde{H} \equiv i\sigma^2 H^*$, $h_t \equiv y_t s_\beta$, $X_t \equiv A_t - \mu \cot\beta$, 
and the hypercharges are $Y_{\tilde{Q}} = 1/6, Y_{\tilde{t}_R} = -2/3$. The mass matrix entries $m_{\tilde{Q}}$ and $m_{\tilde{t}_R}$ are the soft supersymmetry-breaking masses in the MSSM Lagrangian. We note that 
$\tilde{Q} = (\tilde{t}_L \, , \tilde{b}_L)$ is an $SU(2)_L$ doublet, so $U$ is implicitly a $3 \times 3$ matrix,
and there will be an additional trace over color. Substituting this into the CDE expansion
with $G_{\mu\nu}$ the gluon field strength, we extract from the universal one-loop effective action the term 
\begin{equation*}
\mathcal{L}_\text{1-loop}^\text{eff} \supset \frac{1}{(4\pi)^2}\frac{1}{24}\left(\frac{h_t^2 - \frac{1}{6}g_1^2 c_{2\beta}}{m^2_{\tilde{Q}}} + \frac{h_t^2 + \frac{1}{3} g_1^2 c_{2\beta}}{m^2_{\tilde{t}_R}} - \frac{h_t^2 X_t^2}{m^2_{\tilde{Q}} m^2_{\tilde{t}_R} } \right) g_3^2 |H|^2 G^a_{\mu\nu} {G^a}^{\mu\nu}	\, .
\end{equation*}
This yields the dimension-6 operator $\mathcal{O}_g$ in the ESM:
\begin{equation*}
\mathcal{L}_\text{dim-6} \supset \frac{\bar{c}_g}{m_W^2}\mathcal{O}_g 	\, ,
\end{equation*}
with the Wilson coefficient given in this normalisation~\footnote{In general, barred coefficients are related to unbarred ones by $\bar{c} \equiv c \frac{M^2}{\Lambda^2}$ where $M = v, m_W$ depending on the operator normalisation in the Lagrangian.} by 
\begin{equation*}
\bar{c}_g =  \frac{m_W^2}{(4\pi)^2}\frac{1}{24}\left(\frac{h_t^2 - \frac{1}{6}g_1^2 c_{2\beta}}{m^2_{\tilde{Q}}} + \frac{h_t^2 + \frac{1}{3} g_1^2 c_{2\beta}}{m^2_{\tilde{t}_R}} - \frac{h_t^2 X_t^2}{m^2_{\tilde{Q}} m^2_{\tilde{t}_R} } \right)	\, .
\end{equation*}

This example demonstrates the relative ease with which one may obtain a Wilson coefficient at the one-loop level
without having to compute Feynman diagrams in both the UV model and the EFT that then have to be matched, 
a process that must be redone every time one adds a new particle to integrate out. 
Here we may add a right-handed sbottom simply by enlarging the $U$ matrix for
$\Phi = (\tilde{Q} \,, \tilde{t}^*_R \,, \tilde{b}^*_R)$ and plugging it back into (\ref{eq:universalGGdiag}), giving the result 
\begin{equation}
\bar{c}_g =
 \frac{m_W^2}{(4\pi)^2}\frac{1}{24} \left( \frac{h_b^2 + h_t^2 - \frac{1}{6} g_1^2 c_{2\beta}}{ m_{\tilde{Q}^2}} + \frac{h_t^2 + \frac{1}{3} g_1^2 c_{2\beta}}{ m_{\tilde{t}_R^2}}  + \frac{h_b^2 - \frac{1}{6} g_1^2 c_{2\beta}}{ m_{\tilde{b}_R^2}}
- 
\frac{h_t^2 X_t^2}{m_{\tilde{Q}^2}   m_{\tilde{t}_R^2} } - \frac{ h_b^2 X_b^2 }{ m_{\tilde{Q}^2} m_{\tilde{b}_R^2} }  \right)	\, .
\label{eq:cg}
\end{equation}

We compute similarly the dimension-6 operators affecting the $h\gamma\gamma$ coupling,
with the field strength matrix given in this case by
\begin{equation*}
G^\prime_{\mu\nu} = \twomatrix{
{W^\prime}^a_{\mu\nu}\tau^a + Y_{\tilde{Q}}B^\prime_{\mu\nu}\identity & 0 \\
0 & -Y_{\tilde{t}_R}B^\prime_{\mu\nu}
}	\, .
\end{equation*}
Evaluating this in the CDE method then yields directly
\begin{equation*}
\mathcal{L}_\text{dim-6} \supset \frac{\bar{c}_{BB}}{m_W^2}\mathcal{O}_{BB} + \frac{\bar{c}_{WW}}{m_W^2}\mathcal{O}_{WW} + \frac{\bar{c}_{WB}}{m_W^2}\mathcal{O}_{WB}	\, , 
\end{equation*}
where
\begin{equation*}
\mathcal{O}_{BB} = g_1^2 |H|^2 B_{\mu\nu}B^{\mu\nu} \quad , \quad 
\mathcal{O}_{WW} = g_2^2 |H|^2 W^a_{\mu\nu}{W^a}^{\mu\nu}	\quad , \quad
\mathcal{O}_{WB} = 2g_1g_2 H^\dagger\tau^a H W^a_{\mu\nu}B^{\mu\nu}	\, ,
\end{equation*}
and 
\begin{align}
%\begin{eqnarray}
&\frac{(4\pi)^2}{m_W^2} \bar{c}_{BB}= \frac{1}{864} \left(\frac{6 h_t^2-g_{1}^2 c_{2 \beta }}{m_{\tilde{Q}}^2}+\frac{32 \left(g_{1}^2 c_{2 \beta }+3 h_t^2\right)}{m_{\tilde{t}_R}^2}\right) \nonumber \\
&  +h_{t}^2 X_t^2 \left(-\frac{-103 m_{\tilde{Q}}^6 m_{\tilde{t}_R}^2-39 m_{\tilde{Q}}^4 m_{\tilde{t}_R}^4+17 m_{\tilde{Q}}^2 m_{\tilde{t}_R}^6+16 m_{\tilde{Q}}^8+m_{\tilde{t}_R}^8}{144 m_{\tilde{Q}}^2 m_{\tilde{t}_R}^2 \left(m_{\tilde{Q}}^2-m_{\tilde{t}_R}^2\right){}^4} + \frac{\left(m_{\tilde{Q}}^2 m_{\tilde{t}_R}^4-4 m_{\tilde{Q}}^4 m_{\tilde{t}_R}^2\right) \ln \left(\frac{m_{\tilde{Q}}^2}{m_{\tilde{t}_R}^2}\right)}{4 \left(m_{\tilde{Q}}^2-m_{\tilde{t}_R}^2\right){}^5} \right) \, ,  \\
&\frac{(4\pi)^2}{m_W^2} \bar{c}_{WW}= \frac{6 h_t^2-g_{1}^2 c_{2 \beta }}{96 m_{\tilde{Q}}^2}+h_t^2 X_t^2 \left(-\frac{\left(m_{\tilde{Q}}^2+m_{\tilde{t}_R}^2\right) \left(-8 m_{\tilde{Q}}^2 m_{\tilde{t}_R}^2+m_{\tilde{Q}}^4+m_{\tilde{t}_R}^4\right)}{16 m_{\tilde{Q}}^2 \left(m_{\tilde{Q}}^2-m_{\tilde{t}_R}^2\right){}^4}-\frac{3 m_{\tilde{Q}}^2 m_{\tilde{t}_R}^4 \ln \left(\frac{m_{\tilde{Q}}^2}{m_{\tilde{t}_R}^2}\right)}{4 \left(m_{\tilde{Q}}^2-m_{\tilde{t}_R}^2\right){}^5}\right) \, ,  \\
&\frac{(4\pi)^2}{m_W^2} \bar{c}_{WB} =  -\frac{g_{2}^2 c_{2\beta}+2 h_{t}^2}{48 m_{\tilde{Q}}^2}  + h_t^2 X_t^2 \left(\frac{33 m_{\tilde{Q}}^4 m_{\tilde{t}_R}^2-3 m_{\tilde{Q}}^2 m_{\tilde{t}_R}^4+5 m_{\tilde{Q}}^6+m_{\tilde{t}_R}^6}{24 m_{\tilde{Q}}^2 \left(m_{\tilde{Q}}^2-m_{\tilde{t}_R}^2\right){}^4}-\frac{m_{\tilde{Q}}^2 m_{\tilde{t}_R}^2 \left(2 m_{\tilde{Q}}^2+m_{\tilde{t}_R}^2\right) \ln \left(\frac{m_{\tilde{Q}}^2}{m_{\tilde{t}_R}^2}\right)}{2 \left(m_{\tilde{Q}}^2-m_{\tilde{t}_R}^2\right){}^5}\right ) \, . 
\label{eq:cgamma}
%\end{eqnarray}
\end{align}
In the basis used in~\cite{ESY}, the operators $\mathcal{O}_{WW}$ and $\mathcal{O}_{WB}$ are eliminated and 
constraints are placed on $\mathcal{O}_\gamma \equiv \mathcal{O}_{BB}$. 
The coefficients are related by $\bar{c}_\gamma = \bar{c}_{BB} + \bar{c}_{WW} - \bar{c}_{WB}$~\footnote{The log terms in Eqs.~[2.5-2.7] cancel in the conribution to $\bar{c}_\gamma$.}.

To summarise, one may calculate $\bar{c}_g$ and $\bar{c}_\gamma$ from integrating out a heavy complex scalar $\Phi$ in an arbitrary UV model by substituting the SM field matrix, $U(x)$, and field strength matrix, $G_{\mu\nu}$, into the CDE expansion. The computation of one-loop Wilson coefficients is thus reduced to evaluating the trace of a few matrices. These universal results are extendable to all dimension-6 operators and apply also when integrating out heavy fermions and massive or massless gauge bosons~\cite{HLM2, DEQY2}.

\section{Feynman Diagram Calculations and Comparison}

\begin{figure}[h!]
\centering
\vspace*{-2.2cm}
\includegraphics[scale=0.24]{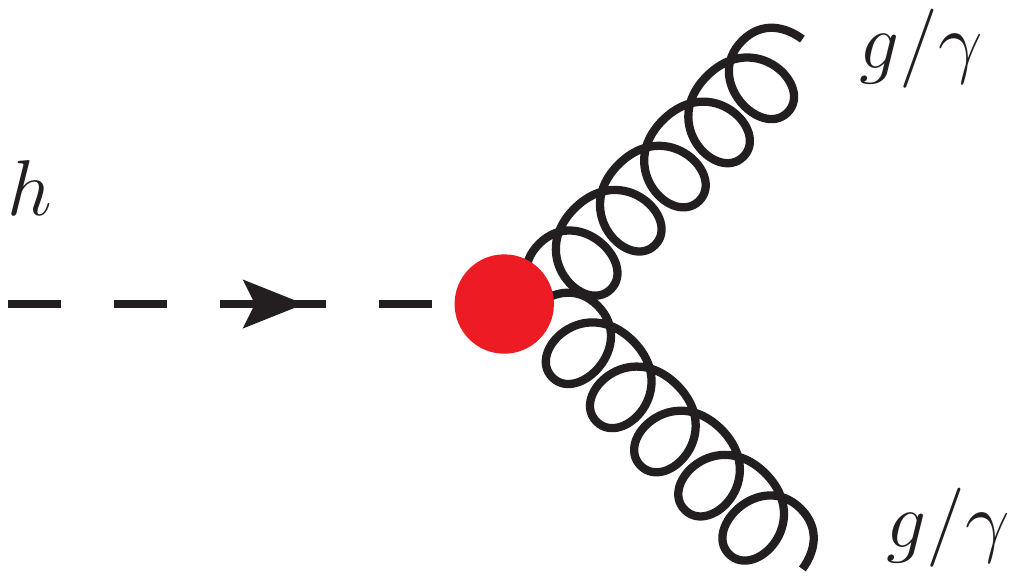}
\includegraphics[scale=0.24]{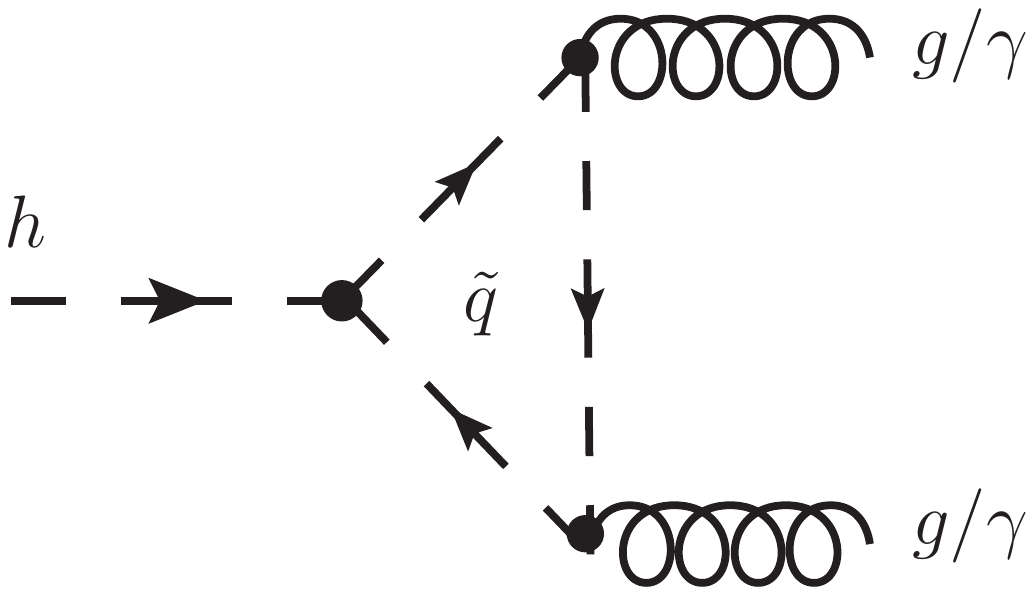}
\includegraphics[scale=0.24]{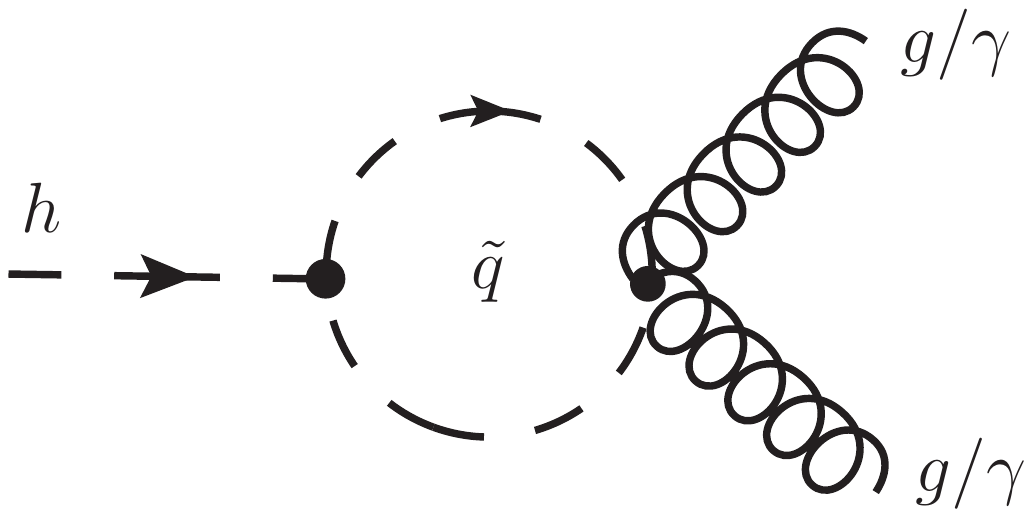}
\vspace*{-2.2cm}
\caption{\it Leading order tree-level Feynman diagram for the EFT (left) and one-loop diagrams for the squark contributions (middle and right) to the $h\to gg/\gamma\gamma$ amplitude. }
\label{fig:feynmandiagrams}
\end{figure}

To estimate quantitatively the validity of the dimension-6 EFT we compare the coefficients obtained above with
results from an exact one-loop calculation in the MSSM.
This is achieved by calculating the Feynman diagrams in Fig.~\ref{fig:feynmandiagrams} then matching the $h \to gg$ and $h \to \gamma\gamma$ amplitudes in the EFT with the equivalent MSSM amplitude. In the EFT the operators $\mathcal{O}_g$ and 
$\mathcal{O}_\gamma$ can be expanded after electroweak symmetry breaking (EWSB) around the 
vacuum expectation value $v \sim 174$ GeV in order to get the Lagrangian 
\begin{equation*}
\mathcal{L}_{hVV} = g_3^{2}\sqrt{2}v \frac{\bar{c}_g}{m_W^{2}} hG^{a}_{\mu\nu}G^{a,\mu\nu}  
+g_1^2\sqrt{2}v \frac{\bar{c}_{\gamma}}{m_W^{2}} hB_{\mu\nu}B^{\mu\nu}	\, ,
\end{equation*}
corresponding to the following Feynman rules for the $hgg$ and $h\gamma\gamma$ vertices:
\begin{align*}
iV_{hgg}^{\mu\nu}(p_2, p_3) &= -4i g_{3}^2 \sqrt{2} v \frac{\bar{c}_g}{m_W^{2}} \big(p_2 p_3 
g^{\mu\nu} - p_2^\nu p_3^\mu \big) \, , \\
iV_{h\gamma\gamma}^{\mu\nu}(p_2, p_3) &= -4i e^{2} \sqrt{2} v \frac{\bar{c}_{\gamma}}{m_W^{2}} 
\big(p_2 p_3 g^{\mu\nu} - p_2^\nu p_3^\mu \big)	\, .
\end{align*}
Thus the $h \to gg$ and $h \to \gamma\gamma$ amplitudes for on-shell external particles are 
\begin{align}
{\cal A}_{EFT}^{hgg} &= -16 g_{s}^{2} \sqrt{2}v \frac{\bar{c}_g}{m_W^{2}} \left( \xi_{2}^{*}.\xi_{3}^{*} M_{h}^2 - 2(\xi_{2}^{*}.p_{1})(\xi_{3}^{*}.p_{1})  \right)	\, ,  
\label{eq:amplitudegg} \\
{\cal A}_{EFT}^{h\gamma\gamma} &= -2 g_1^2 \cos^{2}\theta_{W} \sqrt{2}v \frac{\bar{c}_{\gamma}}{m_W^{2}} \left( \xi_{2}^{*}.\xi_{3}^{*} M_{h}^2 - 2(\xi_{2}^{*}.p_{1})(\xi_{3}^{*}.p_{1}) \right) \, , 
\label{eq:amplitudegaga}
\end{align}
where the $\xi_i$ are the polarization vectors of the gauge bosons.

We computed the one-loop diagrams in Fig.~\ref{fig:feynmandiagrams} in the MSSM
and checked our results using the {\tt FeynArts} package~\cite{feynarts}. The CP-even Higgs bosons are rotated to their physical basis by a mixing angle $\alpha$ which we set to be $\alpha=\beta-\pi/2$ corresponding to the decoupling limit when the pseudo-scalar Higgs mass is much heavier than the mass of the Z gauge boson, as indicated by the experimental data~\cite{hMSSM} and appropriate to our scenario of light stops~\footnote{The case of relatively heavy stops has been demonstrated to be described in a very compact and convenient way, depending only on the two parameters $\tan{\beta}$ and the pseudo-scalar Higgs mass, when the observed Higgs mass is taken into account~\cite{hMSSM}.}.

When comparing the EFT and MSSM amplitudes we may choose the momenta of the 
external particles to be on-shell for convenience. The result of this procedure for the $h \to gg$ amplitude
yields the same expression as (\ref{eq:amplitudegg}) with the replacement $\bar{c}_g \to \bar{c}^\text{MSSM}_g$, where
\begin{equation}
\bar{c}^\text{MSSM}_g = (\bar{c}^\text{MSSM}_g)^{\tilde{t}} + (\bar{c}^\text{MSSM}_g)^{\tilde{b}}	\, ,
\label{eq:exactcg}
\end{equation}
where the part due to stops is given by
\begin{align*}
& \quad\quad (\bar{c}^\text{MSSM}_g)^{\tilde{t}} = \frac{ m_W^2}{ 6 (4\pi)^{2}}  \frac{  N^{\tilde{t}}_g    }{  D^{\tilde{t}}_g   }	 \, ,  \\
&N^{\tilde{t}}_g =  \frac{c_{2\beta} g_1^2}{s^{2}_{W}} \left[v^2 c_{2\beta} g_1^2 \left(2 c_{2W} +1\right)+3 \left(3 v^2 h_t^2+2 \left(m_{\tilde{t}_R}^2-m_{\tilde{Q}}^2\right) c_{2W}+2 m_{\tilde{Q}}^2+m_{\tilde{t}_R}^2\right)\right] \\ 
& \quad\quad + 36 h_t^2 \left(v^2 h_t^2+m_{\tilde{Q}}^2+m_{\tilde{t}_R}^2-X_t^2\right) 	\, , \\
&D^{\tilde{t}}_g =    \frac{v^2 c_{2\beta} g_1^2}{s^{2}_{W}} \left[v^2 c_{2\beta} g_1^2 \left(2 c_{2W}+1\right)+3 \left(3 v^2 h_t^2+4 \left(m_{\tilde{t}_R}^2-m_{\tilde{Q}}^2\right) c_{2W}+4 m_{\tilde{Q}}^2+2 m_{\tilde{t}_R}^2\right)\right]  \\
& \quad\quad +36 \left(v^2 h_t^2+2 m_{\tilde{Q}}^2\right) \left(v^2 h_t^2+2 m_{\tilde{t}_R}^2\right)-72 v^2 h_t^2 X_t^2  \, ,
\end{align*}
and the sbottom contribution reads, 
\begin{equation*}
(\bar{c}^\text{MSSM}_g)^{\tilde{b}} =\frac{m_W^2}{6 (4\pi)^{2}} \frac{c_{2\beta} g_1^2 \left\{6 \left[\left(m_{\tilde{b}_R}^2-m_{\tilde{Q}}^2\right) c_{2W}+m_{\tilde{Q}}^2+2 m_{\tilde{b}_R}^2\right]-v^2 c_{2\beta} g_1^2 \left(c_{2W}+2\right)\right\}}{ \left(12 m_{\tilde{b}_R}^2-v^2 c_{2\beta} g_1^2\right) \left[v^2 c_{2\beta} g_1^2 \left(c_{2W}+2\right)-24m_{\tilde{Q}}^2 s^{2}_{W}\right]}	\, .
\end{equation*}
For $\bar{c}_\gamma$ we simply have
\begin{equation}
\bar{c}^\text{MSSM}_\gamma =  \frac{8}{3}(\bar{c}^\text{MSSM}_g)^{\tilde{t}}  +  \frac{3}{2}(\bar{c}^\text{MSSM}_g)^{\tilde{b}}	\, .
\label{eq:exactcgamma}
\end{equation}
In the limit $v\to 0$ we obtain the same expressions as $\bar{c}_g$ and $\bar{c}_\gamma$ in (\ref{eq:cg})
and (\ref{eq:cgamma}), respectively. Since $\bar{c}_g$ and $\bar{c}_\gamma$ correspond to 
a truncation of the full theory at the dimension-6 level, they contain only the leading-order
terms in an expansion in inverse powers of the stop mass, 
whereas the MSSM result is exact and include higher-order terms in $v/m_{\tilde{t},\tilde{b}}$ that would be generated
by higher-dimensional operators in the EFT approach. Therefore,
we expect the discrepancy between the two approaches to scale with the 
ratio $v/m_{\tilde{t},\tilde{b}}$ for  $m_{\tilde{t},\tilde{b}}$, and the differences between the EFT and exact MSSM results gives
insight into the potential importance of such higher-dimensional operators. 
We note that a large value of $X_t$ in terms like $v^2 m_W^2 X_t^2 / m^6_{\tilde{t}}$ could potentially affect the 
validity of the EFT even for large stop masses, but the positivity of the lightest physical mass eigenvalue 
imposes an upper limit $X_t \simeq m_{\tilde{t}}^2 / m_t$. 

\begin{figure}[h!]
\centering
\hspace*{-0.4cm}
\includegraphics[scale=0.4]{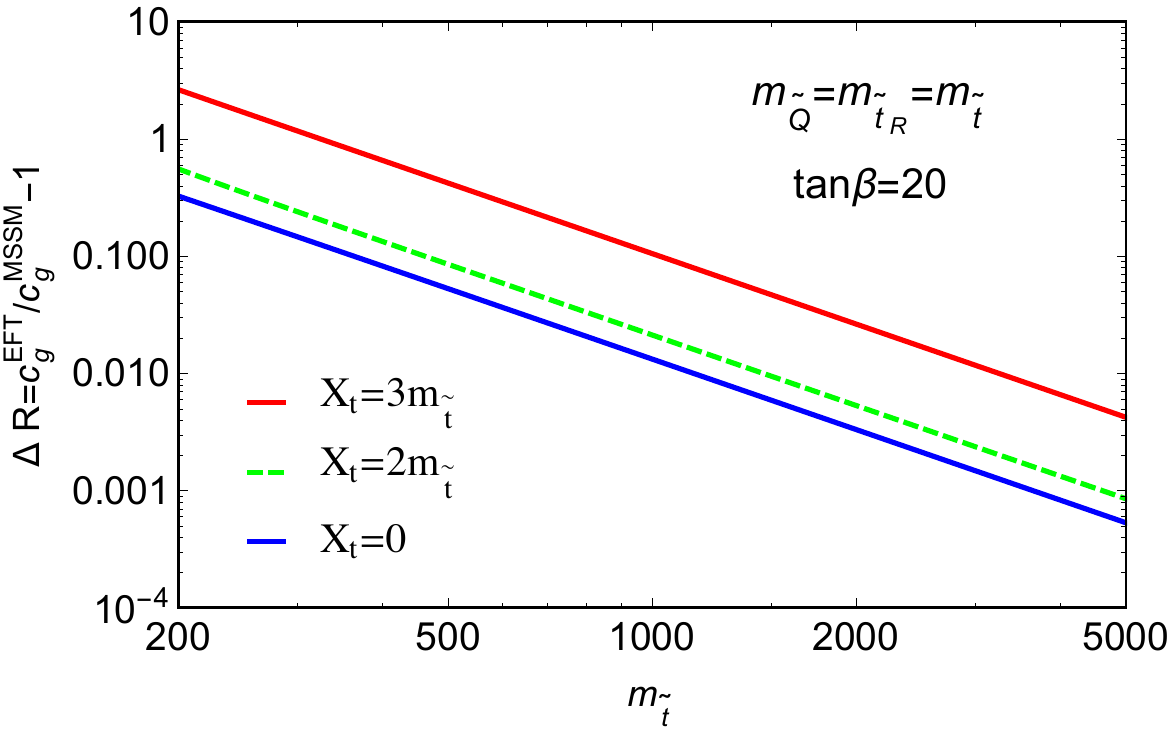}
\includegraphics[scale=0.4]{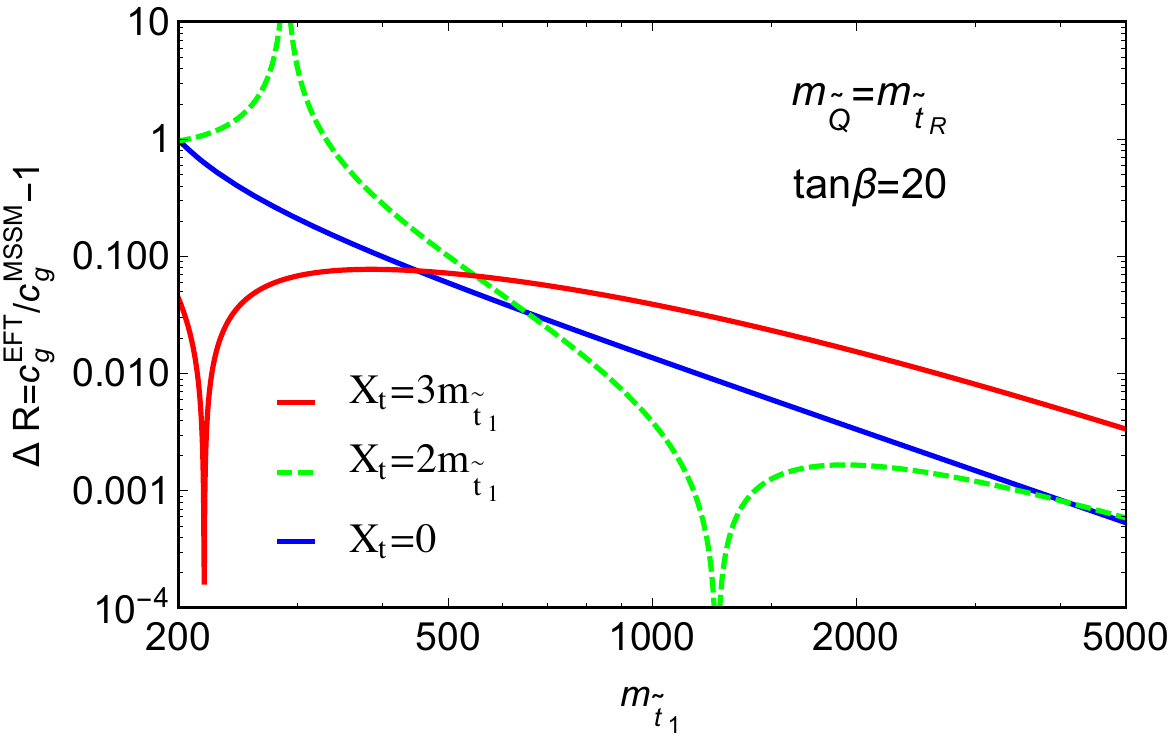}
\caption{\it Values of $\Delta R$, defined in (\protect\ref{DeltaR}), in the degenerate case
$m_{\tilde Q} = m_{\tilde t_R} \equiv m_{\tilde t}$ for $\tan \beta = 20$ and the indicated values of $X_t$,
as a function of $m_{\tilde t}$ (left panel), and as functions of $m_{\tilde t_1}$ (right panel).}
\label{fig:DeltaR}
\end{figure}

The physical mass eigenstates are obtained by diagonalizing the squark mass matrices~\cite{Djouadi:2005gj}
\bear 
\label{matrix_sqmass}
{\cal M}^2_{\tilde{q}} =
\left(
  \begin{array}{cc} m_q^2 + m_{LL}^2 & m_q \, X_q  \\
                    m_q\, X_q    & m_q^2 + m_{RR}^2 
  \end{array} \right) 
\eear
with the various entries defined by
\begin{align}
m_{LL}^2 &= m_{\tilde{Q}}^2 + (I^{3L}_q - Q_q s_W^2)\, M_Z^2\, c_{2\beta} \, , \\
m_{RR}^2 &= m_{\tilde{q}_R}^2 + Q_q s_W^2\, M_Z^2\, c_{2\beta} \, , \\
 X_q  &= A_q - \mu (\tb)^{-2 I_q^{3L}} \, .
\label{mass-matrix}
\end{align}
$Q_q$ and $I^{3L}_q$ is the electromagnetic charge and the weak doublet isospin respectively. After rotating the $2 \times 2$ matrices by an angle $\theta_q$,  
which transforms the interaction eigenstates $\tilde{q}_L$ and $\tilde{q}_R$ into
the mass eigenstates $\tilde{q}_1$ and $\tilde{q}_2$, the mixing angle and physical squark masses are given by 
\bear
s_{2\theta_q} = \frac{2 m_q X_q} { m_{\tilde{q}_1}^2
-m_{\tilde{q}_2}^2 } \ \ , \ \ 
c_{2\theta_q} = \frac{m_{LL}^2 -m_{RR}^2} 
{m_{\tilde{q}_1}^2 -m_{\tilde{q}_2}^2 } \hspace*{1.8cm} \\  
m_{\tilde{q}_{1,2}}^2 = m_q^2 +\frac{1}{2} \left[
m_{LL}^2 +m_{RR}^2 \mp \sqrt{ (m_{LL}^2
-m_{RR}^2 )^2 +4 m_q^2 X_q^2 } \ \right] \, .
\label{stop_mass_eigenvalues}
\eear
We see that in the stop sector the mixing is strong for large values of the parameter
$X_t=A_t- \mu \cot \beta$, which generates a large mass splitting between the two physical mass 
eigenstates and makes $\tilde{q}_1$ much lighter
than the other sparticle $\tilde{q}_2$.

We now compare the values of the $\bar{c}_g$ coefficients calculated in the MSSM and the 
EFT~\footnote{We omit RGE effects that mix the coefficients in the running~\cite{dim6RGE}, as they would
be higher-order corrections beyond the one-loop level of our analysis.}:
\begin{equation}
\Delta R \; \equiv \; \frac{\bar{c}^\text{EFT}_g}{\bar{c}^\text{MSSM}_g} - 1 \, .
\label{DeltaR}
\end{equation}
Fig.~\ref{fig:DeltaR} displays values of $\Delta R$ for the degenerate case
$m_{\tilde Q} = m_{\tilde t_R} \equiv m_{\tilde t}$, three different values of $X_t$
and the representative choice $\tan \beta = 20$. In the left panel we plot $\Delta R$ as functions of
$m_{\tilde t}$, and the right panel shows $\Delta R$ as functions of the lighter stop mass,
$m_{\tilde t_1}$. We see that in both cases $\Delta R \lesssim 0.1$ for $m_{\tilde t} (m_{\tilde t_1})
\gtrsim 500$~GeV, with a couple of exceptions. One is for the relatively large value $X_t = 3 m_{\tilde t}$
in the left panel, for which $\Delta R \gtrsim 0.1$ for $m_{\tilde t} \lesssim 1000$~GeV, and the other
is for $X_t = 2 m_{\tilde t_1}$ and $m_{\tilde t_1} \sim 290$~GeV in the right panel, which is due to
a node in $\bar{c}^\text{MSSM}_g$. These results serve as a warning that, although the EFT
approach is in general quite reliable for stop mass parameters $\gtrsim 500$~GeV, care should always
be exercised for masses $\lesssim 1000$~GeV.

\begin{figure}[h!]
\begin{center} 
%\hspace*{-2.3cm}
\includegraphics[scale=0.12]{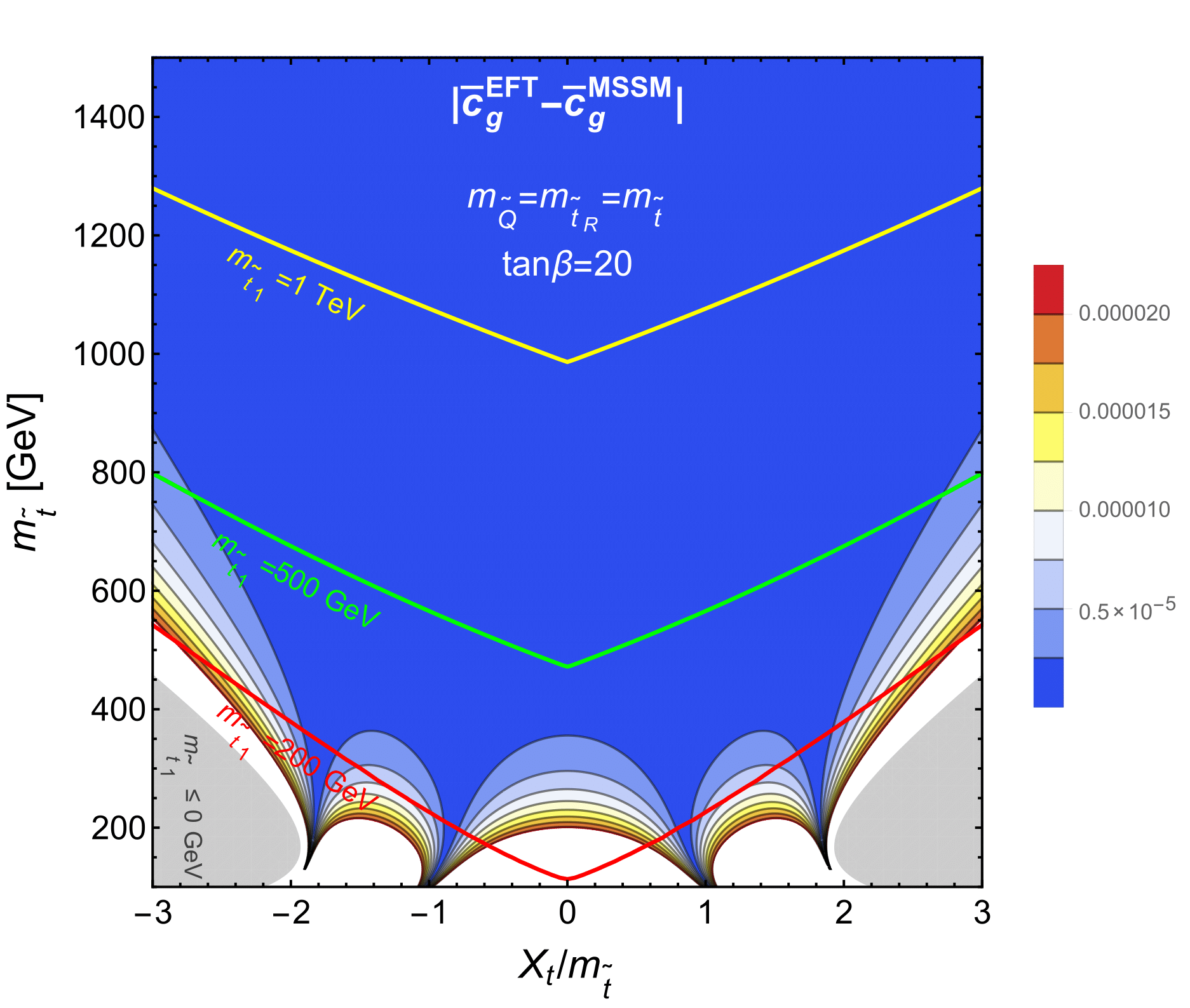}
\includegraphics[scale=0.12]{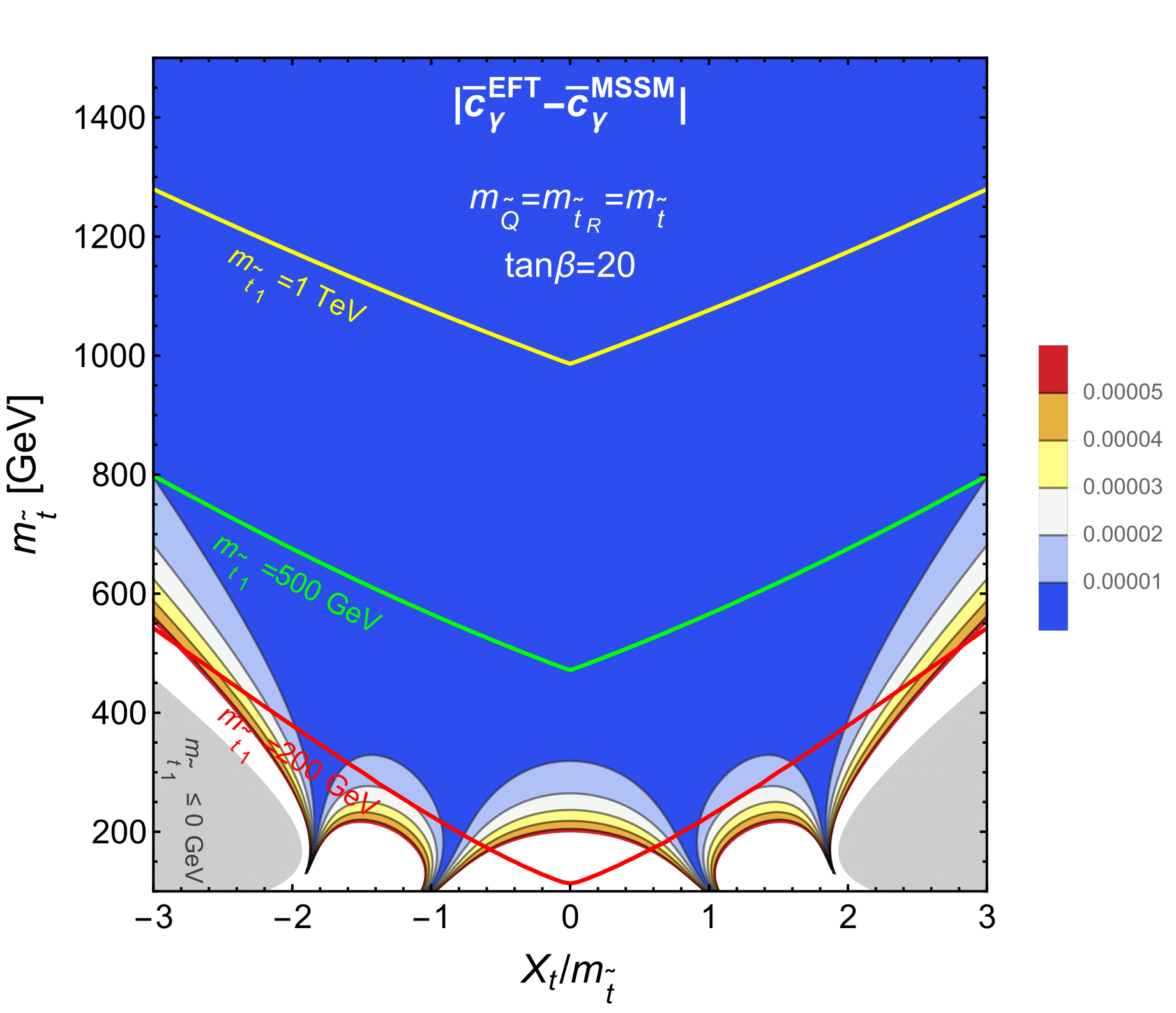}
\caption{\it Contours of the differences $|\bar{c}_g^{\rm EFT} -  \bar{c}_g^{\rm MSSM}|$ (left panel) and 
$|\bar{c}_{\gamma}^{\rm EFT} -  \bar{c}_{\gamma}^{\rm MSSM}|$ (right panel) in $(X_t/m_{\tilde t}, m_{\tilde t})$
planes for the degenerate case $m_{\tilde Q} = m_{\tilde t_R} \equiv m_{\tilde t}$ with $\tan \beta = 20$.
Also shown are contours of $m_{\tilde t_1} = 200$~GeV, $500$~GeV and $1$~TeV and
regions where the ${\tilde t_1}$ becomes tachyonic.}
\label{fig:Delta_ci}
\end{center}
\end{figure}

A similar message is conveyed by Fig.~\ref{fig:Delta_ci}, which uses colour-coding to display
values of the differences $|\bar{c}_g^{\rm EFT} -  \bar{c}_g^{\rm MSSM}|$ (left panel) and 
$|\bar{c}_{\gamma}^{\rm EFT} -  \bar{c}_{\gamma}^{\rm MSSM}|$ (right panel) in $(X_t/m_{\tilde t}, m_{\tilde t})$
planes for the degenerate case $m_{\tilde Q} = m_{\tilde t_R} \equiv m_{\tilde t}$ with $\tan \beta = 20$.
Also shown are contours of $m_{\tilde t_1} = 200$~GeV (red), 500~GeV (green) and 1~TeV (yellow) and
regions where the ${\tilde t_1}$ becomes tachyonic (shaded grey). We see that the differences are
generally $< 2.5 \times 10^{-6}$ for $|\bar{c}_g^{\rm EFT} -  \bar{c}_g^{\rm MSSM}|$ and
$< 10^{-5}$ for $|\bar{c}_{\gamma}^{\rm EFT} -  \bar{c}_{\gamma}^{\rm MSSM}|$ when $m_{\tilde t_1} > 500$~GeV,
even for large values of $X_t$, but that much larger differences are possible for $m_{\tilde t_1} < 200$~GeV,
even for small values of $X_t$.

\section{Constraints on Light Stops from a Global Fit}

\begin{figure}[h!]
\begin{center} 
%\hspace*{-2.3cm}
\includegraphics[scale=0.25]{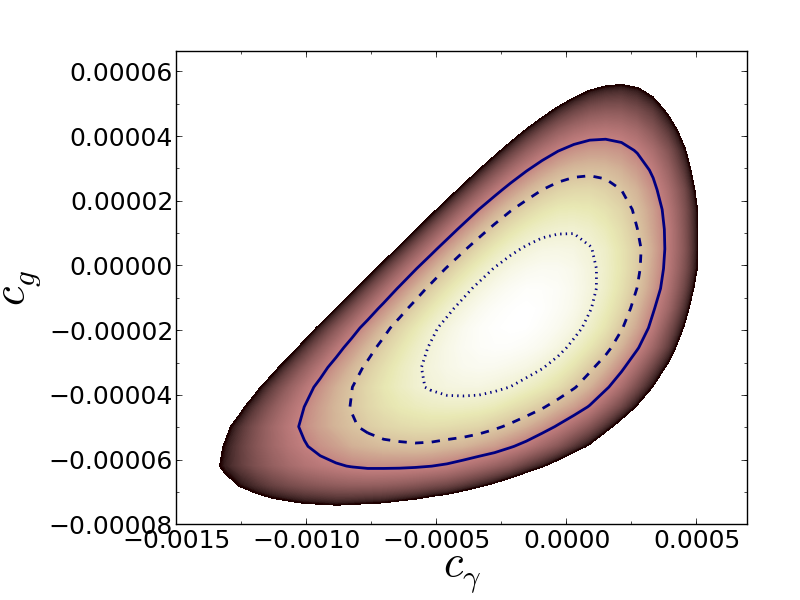}
\includegraphics[scale=0.25]{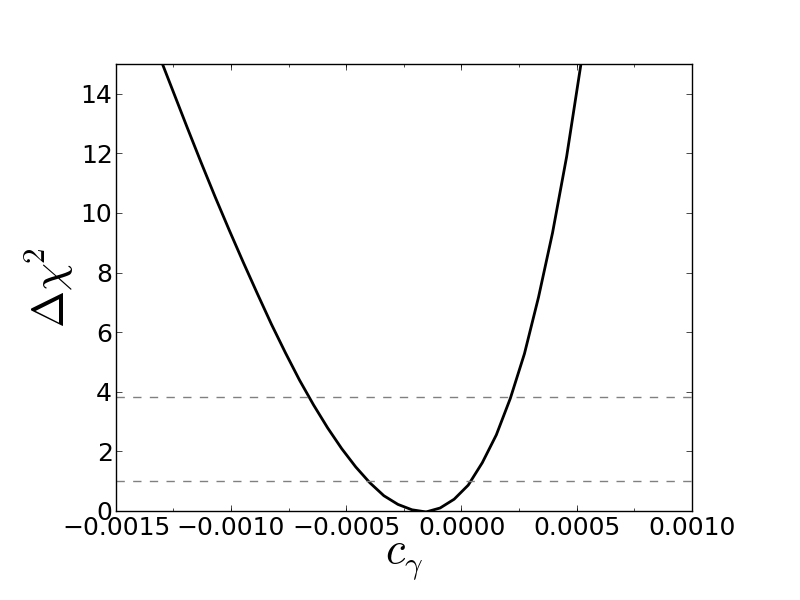}
\includegraphics[scale=0.25]{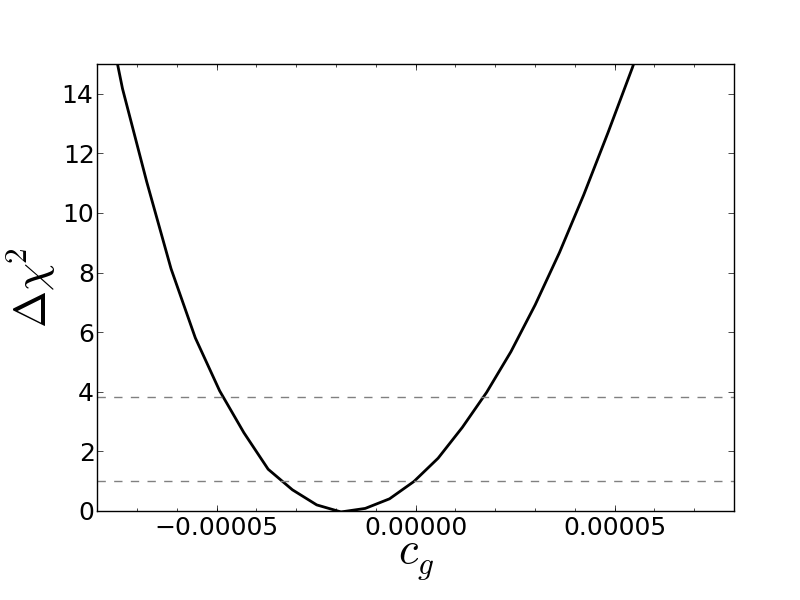}
\caption{\it Results based on the global fit in~\cite{ESY}, varying $\bar{c}_g$ and $\bar{c}_\gamma$ simultaneously
but setting to zero the coefficients of the other dimension-6 operators contributing to the Higgs sector.
The dotted, dashed and solid contours on the left denote the allowed 68\%, 95\% and 99\% CL regions respectively.
The middle and right figures show the marginalized $\chi^2$ functions for $\bar{c}_\gamma$ and $\bar{c}_g$ respectively.   }
\label{fig:cgammavscg}
\end{center}
\end{figure}

We now discuss the constraints on the lighter stop mass that are imposed by the
current experimental constraints on the coefficients $\bar{c}_g$ and $\bar{c}_{\gamma}$,
comparing them with the constraints imposed by electroweak precision observables
via the oblique parameters $S$ and $T$~\cite{PeskinTakeuchi}, as well as the ranges favoured by measurements of
the Higgs mass $M_h$ and direct searches at the LHC.
We note that the $S$ and $T$ parameters are related to the dimension-6 operator coefficients
$\bar{c}_W$, $\bar{c}_B$ and $\bar{c}_T$, as defined in the basis of \cite{ESY}~\footnote{In 
other bases $\bar{c}_W$ and $\bar{c}_B$ may be eliminated in favour of $\bar{c}_{WB}$.}, through
\begin{align*}
S &= \frac{4\sin^2{\theta_W}}{\alpha(m_Z)}(\bar{c}_W+\bar{c}_B) \approx 119 (\bar{c}_W+\bar{c}_B) 	\, , \\
T &= \frac{1}{\alpha(m_Z)} \bar{c}_T \approx 129 \bar{c}_T 	\, .
\end{align*}
We shall quote the electroweak precision constraints on $\bar{c}_W+\bar{c}_B$ and $\bar{c}_T$ instead of $S$ and $T$, 
in keeping with the EFT approach. The stop contributions to these coefficients were given in \cite{HLM1, HLM2},
and Table~\ref{tab:currentconstraints} displays the current experimental constraints on 
$\bar{c}_g, \bar{c}_{\gamma}, \bar{c}_T$ and $\bar{c}_W+\bar{c}_B$ that we apply.

\begin{table}[h]
\begin{center}
\begin{tabular}
{|c | c | c | c  | c |}
\hline
\multirow{2}{*}{Coeff.} &\multicolumn{2}{|c|}{ \multirow{2}{*}{Experimental constraints}} & \multirow{2}{*}{95 \% CL limit} & \multicolumn{1}{|c|}{deg. $m_{\tilde{t}_{1}}$,} \\
 & \multicolumn{2}{|c|}{} & & $X_t=0$ \\
\hline
\multirow{2}{*}{$\bar{c}_g$} & \multirow{2}{*}{LHC} & marginalized & $[-4.5, 2.2] \times 10^{-5}$ & $\sim 410$ GeV \\ 
 & & individual  & $[-3.0,2.5] \times 10^{-5}$ & $\sim 390$ GeV \\
\hline
\multirow{2}{*}{$\bar{c}_\gamma$} & \multirow{2}{*}{LHC} & marginalized  & $[-6.5 ,2.7] \times 10^{-4}$	 & $\sim 215$ GeV \\
 & & individual  & $[-4.0, 2.3] \times 10^{-4}$ & $\sim 230$ GeV \\
\hline
\multirow{2}{*}{$\bar{c}_T$} & \multirow{2}{*}{LEP} & marginalized & $[-10, 10] \times 10^{-4}$ & $\sim 290$ GeV  \\
 & & individual  & $[-5, 5] \times 10^{-4}$ & $\sim 380$ GeV \\
\hline
\multirow{2}{*}{$\bar{c}_W+\bar{c}_B$} & \multirow{2}{*}{LEP} & marginalized & $[-7, 7] \times 10^{-4}$ & $\sim 185$ GeV \\
 & & individual  & $[-5, 5] \times 10^{-4}$ & $\sim 195$ GeV \\
\hline
\end{tabular}
\end{center}
\caption{\it List of the experimental $95\%$ CL bounds on coefficients used in setting current limits on stops,
which are taken from~\protect\cite{ESY}. The marginalized LHC limits are for a two-parameter fit allowing 
$\bar{c}_g$ and $\bar{c}_\gamma$ to vary, and the marginalized LEP limits are for a two-parameter fit of 
$\bar{c}_T$ and $\bar{c}_W + \bar{c}_B$. The corresponding lightest stop mass limits shown are for 
degenerate soft-supersymmetry breaking masses $m_{\tilde{Q}} = m_{\tilde{t}_R} = m_{\tilde{t}}$ with $X_t = 0$. }
\label{tab:currentconstraints}
\end{table}

The constraints on the coefficients in the penultimate column of
Table~\ref{tab:currentconstraints} are taken from a recent
global analysis~\cite{ESY} of LEP, LHC and Tevatron data on Higgs production and
triple-gauge couplings. For $\bar{c}_g$ and $\bar{c}_\gamma$ we list the current 95\% CL 
ranges after marginalising a two-parameter fit in which both
$\bar{c}_g$ and $\bar{c}_{\gamma}$ are allowed to vary~\footnote{ In any specific model 
there may be model-dependent correlations between operator coefficients. In the case with only light stops and nothing else one expects the relation between $\bar{c}_g$ and $\bar{c}_\gamma$ shown in
(\ref{eq:exactcgamma}) to hold, as studied in~\cite{NSUSY}. Here we use the more conservative marginalized ranges
shown in the middle and right panels of Fig.~\ref{fig:cgammavscg},
thereby allowing for additional loop contributions to $\bar{c}_g$ or $\bar{c}_\gamma$. },
as well as considering the more restrictive ranges
found when only $\bar{c}_g$ or $\bar{c}_{\gamma} \ne 0$ individually, with the other operator
coefficients set to zero. Similar marginalized and individual 95\% CL limits on $\bar{c}_T$ and 
$\bar{c}_W+\bar{c}_B$ are displayed, where the two-parameter fit varying $\bar{c}_T$ and 
$\bar{c}_W + \bar{c}_B$ simultaneously is equivalent to the $S, T$ ellipse, as reproduced in \cite{ESY}. 
We note that the stop contributions to the coefficients of the
other relevant operators are far smaller than the ranges of these coefficients that were
found in the global fit. This indicates that one is justified in setting these other
operator coefficients to zero when considering bounds on the stop sector, if one
assumes that there are no important contributions from other possible new physics.

\subsection{Degenerate Stop Masses}

\begin{figure}[h!]
\begin{center} 
%\hspace*{-2.3cm}
\includegraphics[scale=0.38]{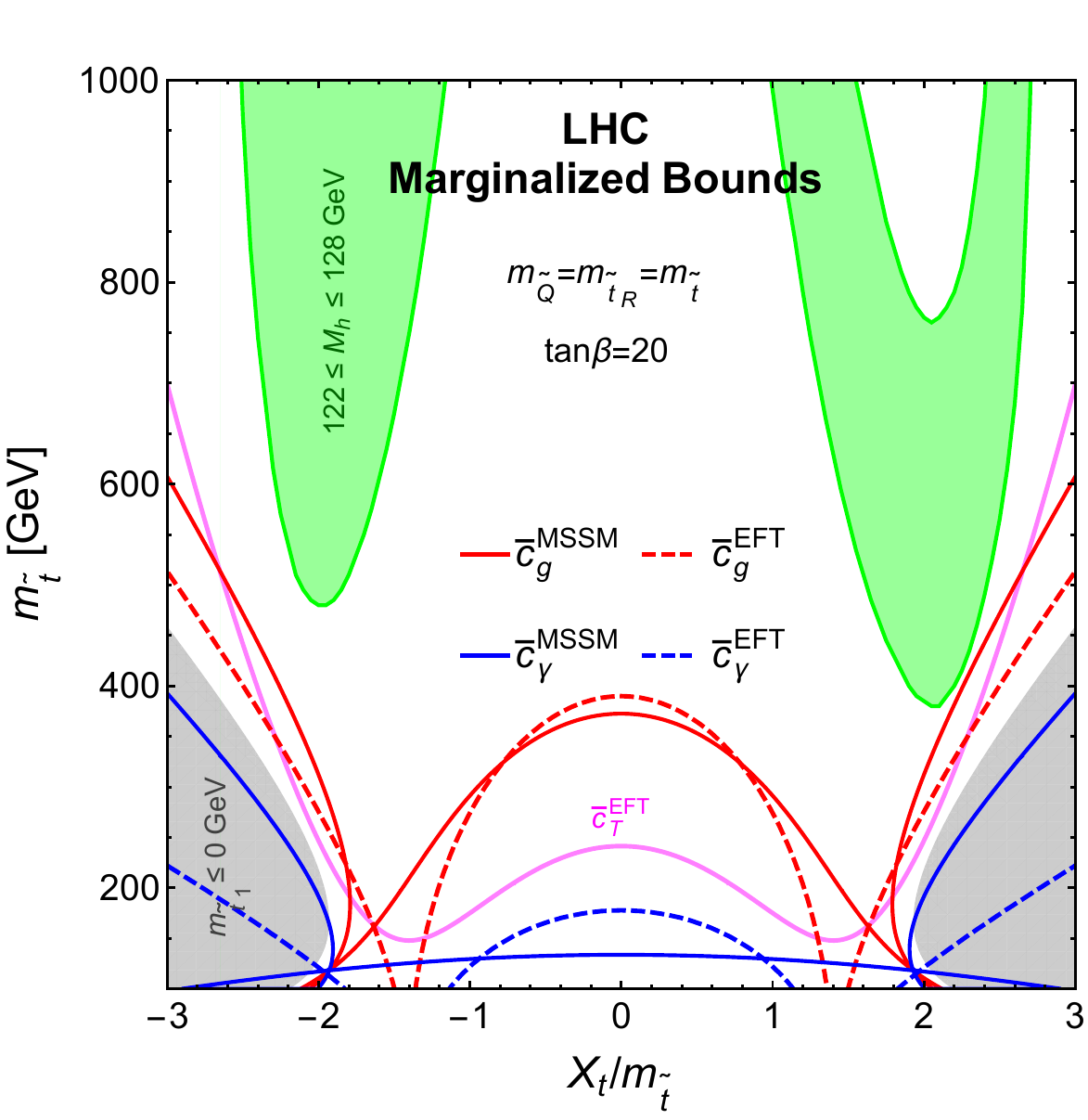}
\includegraphics[scale=0.38]{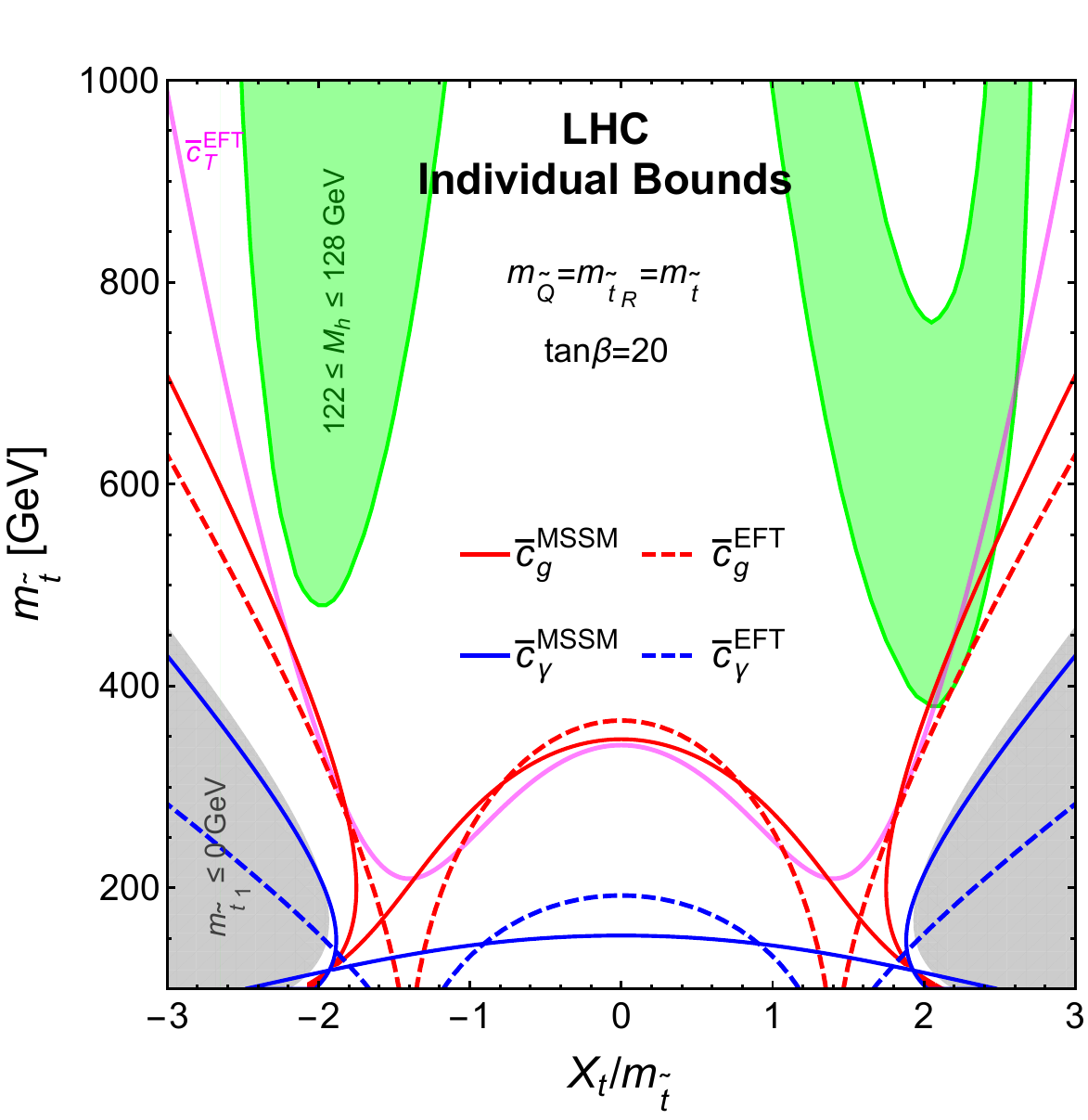}	\\
\includegraphics[scale=0.38]{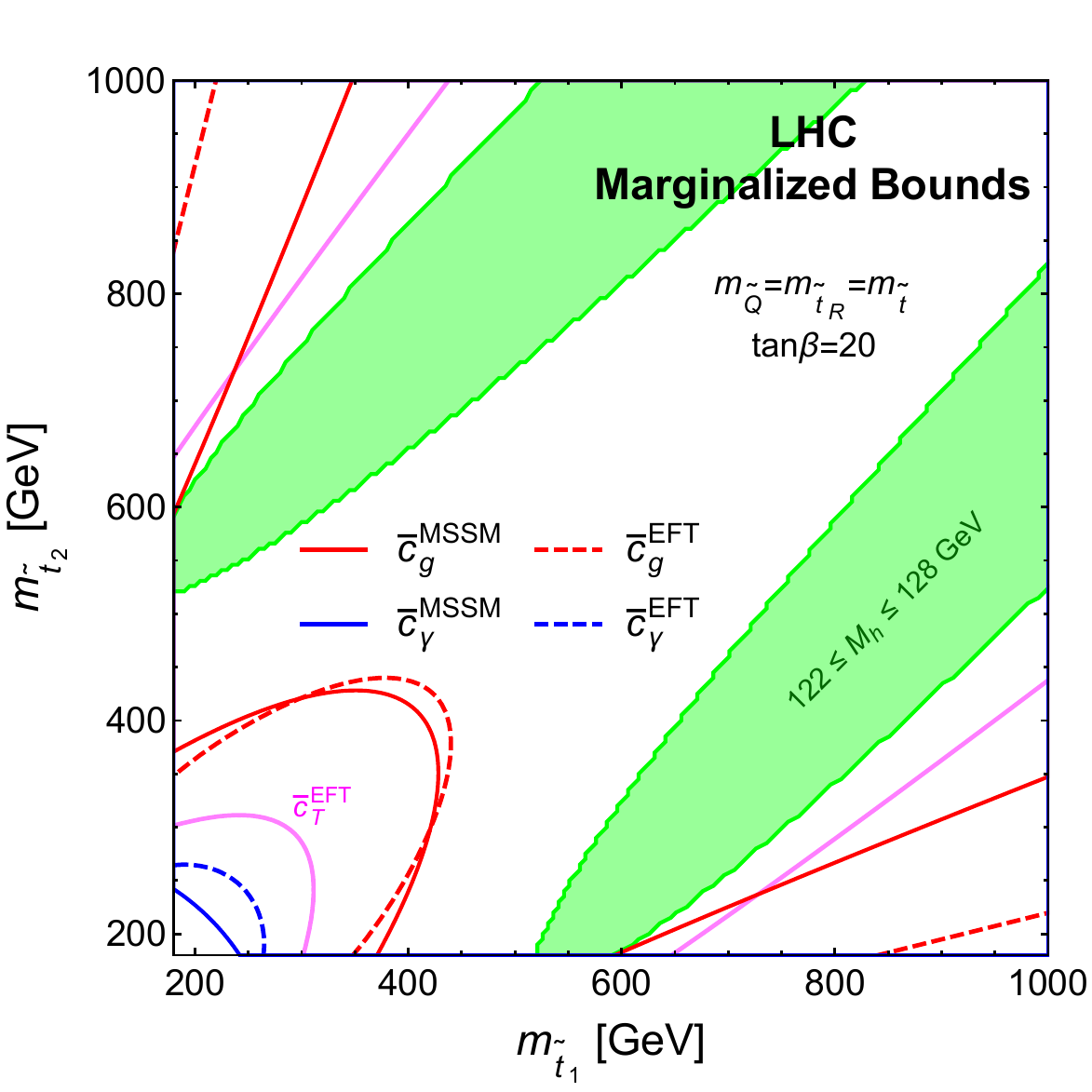}
\includegraphics[scale=0.38]{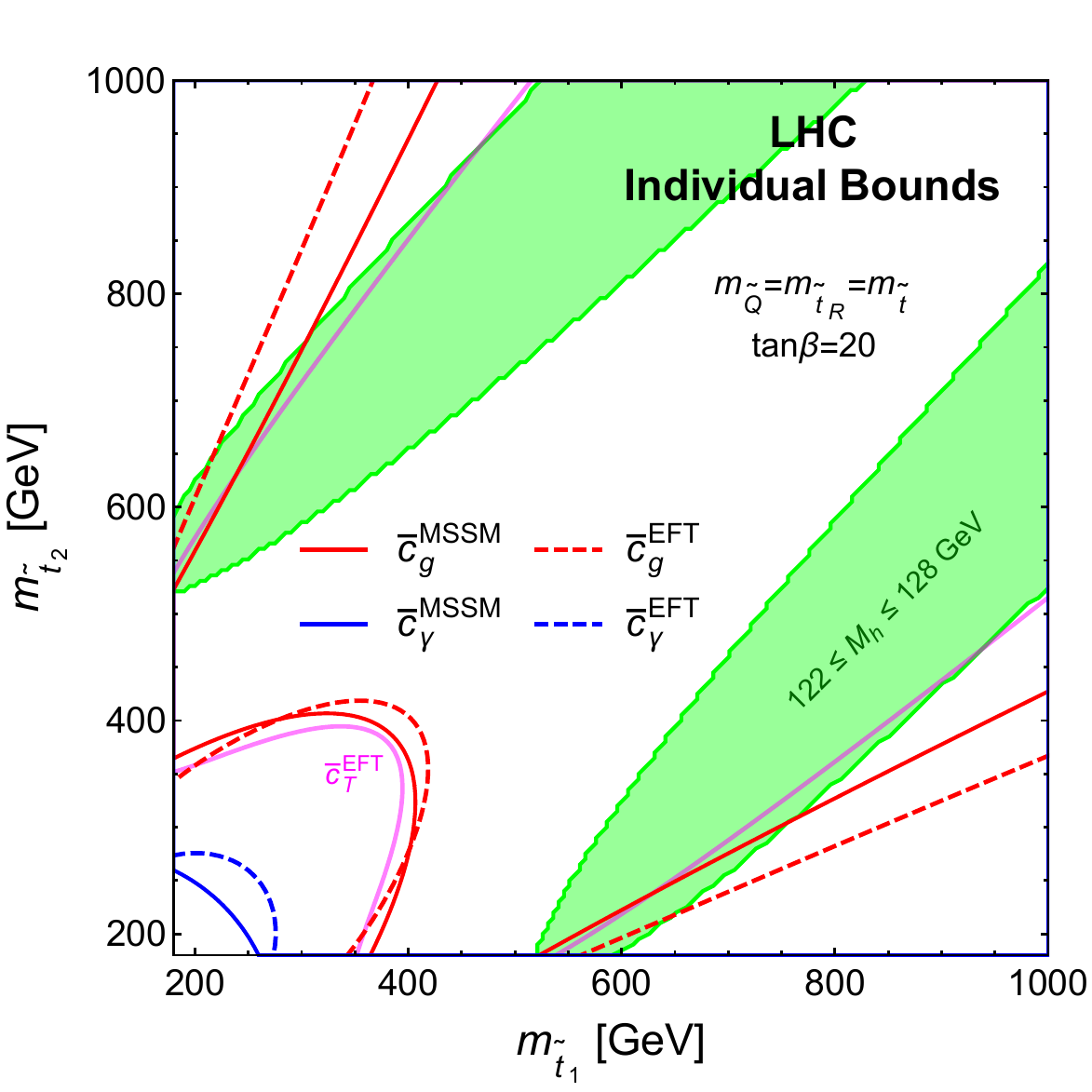}
\caption{\it Compilation of the constraints in (upper panels) the $(X_t/m_{\tilde t_1}, m_{\tilde t})$
plane and (lower panels) the $(m_{\tilde t}, m_{\tilde t_2})$ plane from (left panels) the
marginalized bounds on $\bar{c}_g$ (red lines) and $\bar{c}_{\gamma}$
(blue lines), and from (right panels) the individual bounds on $\bar{c}_g$ and $\bar{c}_{\gamma}$.
Also shown are the EFT bounds on $\bar{c}_T$ (purple lines), the constraint that the lighter
stop should not be tachyonic (grey shading) and the region where $M_h \in (122, 128)$~GeV
according to a {\tt FeynHiggs~2.10.3}~\cite{FH} calculation assuming no other significant contributions
from outside the stop sector (green shading).  }
\label{fig:current_deg_stops}
\end{center}
\end{figure}

Fig.~\ref{fig:current_deg_stops} displays the current constraints in the case of
degenerate soft masses $m_{\tilde Q} = m_{\tilde t_R} \equiv m_{\tilde t}$
with decoupled sbottoms, in the upper panels for $m_{\tilde t}$
as functions of $X_t/m_{\tilde t}$ and in the lower panels for $m_{\tilde t_2}$ as
functions of $m_{\tilde t_1}$, in both cases for $\tan \beta = 20$.
The left panels show the stop constraints from the current marginalized 95\% bounds on $\bar{c}_g$ (red lines)
and $\bar{c}_{\gamma}$ (blue lines), and the right panels show the corresponding bounds from
the current marginalized 95\% bounds. The solid (dashed) lines are obtained from an exact
one-loop MSSM analysis and the EFT approach, respectively.
The purple lines show the individual bound from $\bar{c}_T$ in the EFT approach.
The bounds from $\bar{c}_W + \bar{c}_B$ corresponding to the $S$ parameter are negligible and omitted here.
The grey shaded regions are excluded because the lighter stop becomes tachyonic,
and the green shaded regions correspond to 122~GeV$ < M_h <$128~GeV, as calculated using
{\tt FeynHiggs~2.10.3}~\cite{FH}, allowing for a theoretical uncertainty of $\pm 3$~GeV 
and assuming that there are no other important MSSM
contributions to $M_h$.

\begin{figure}[h!]
\begin{center} 
%\hspace*{-2.3cm}
\includegraphics[scale=0.38]{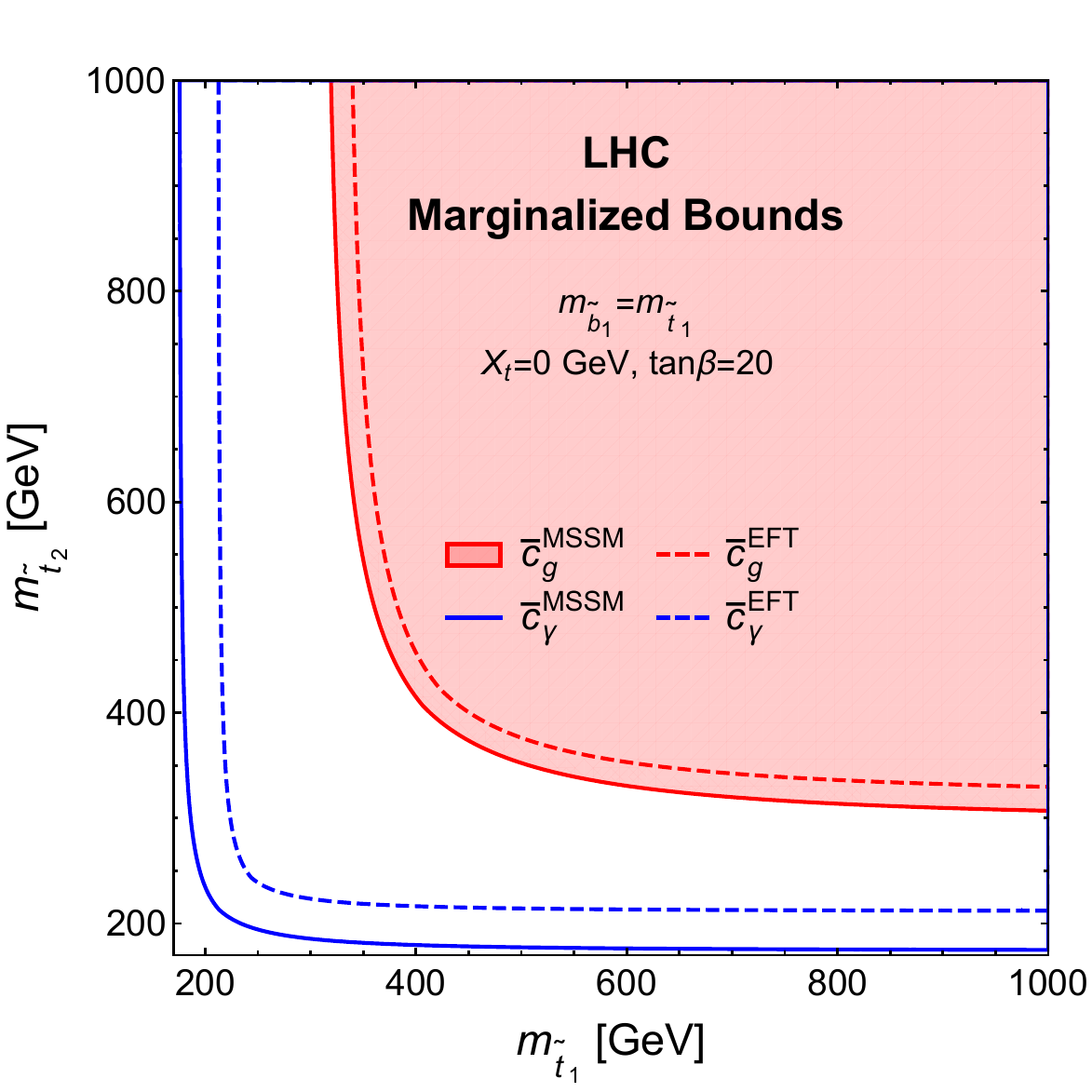}
\includegraphics[scale=0.38]{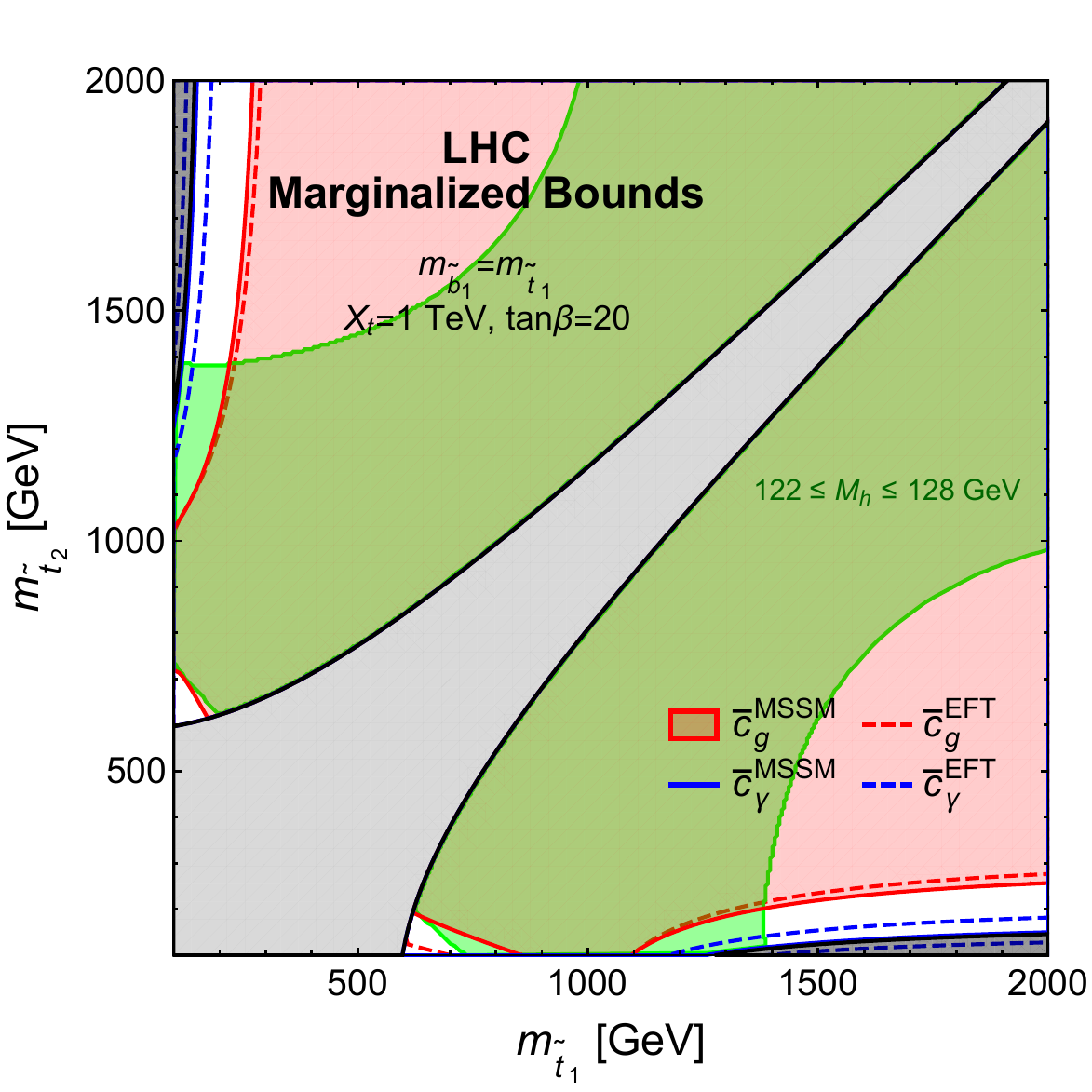}\\
\includegraphics[scale=0.38]{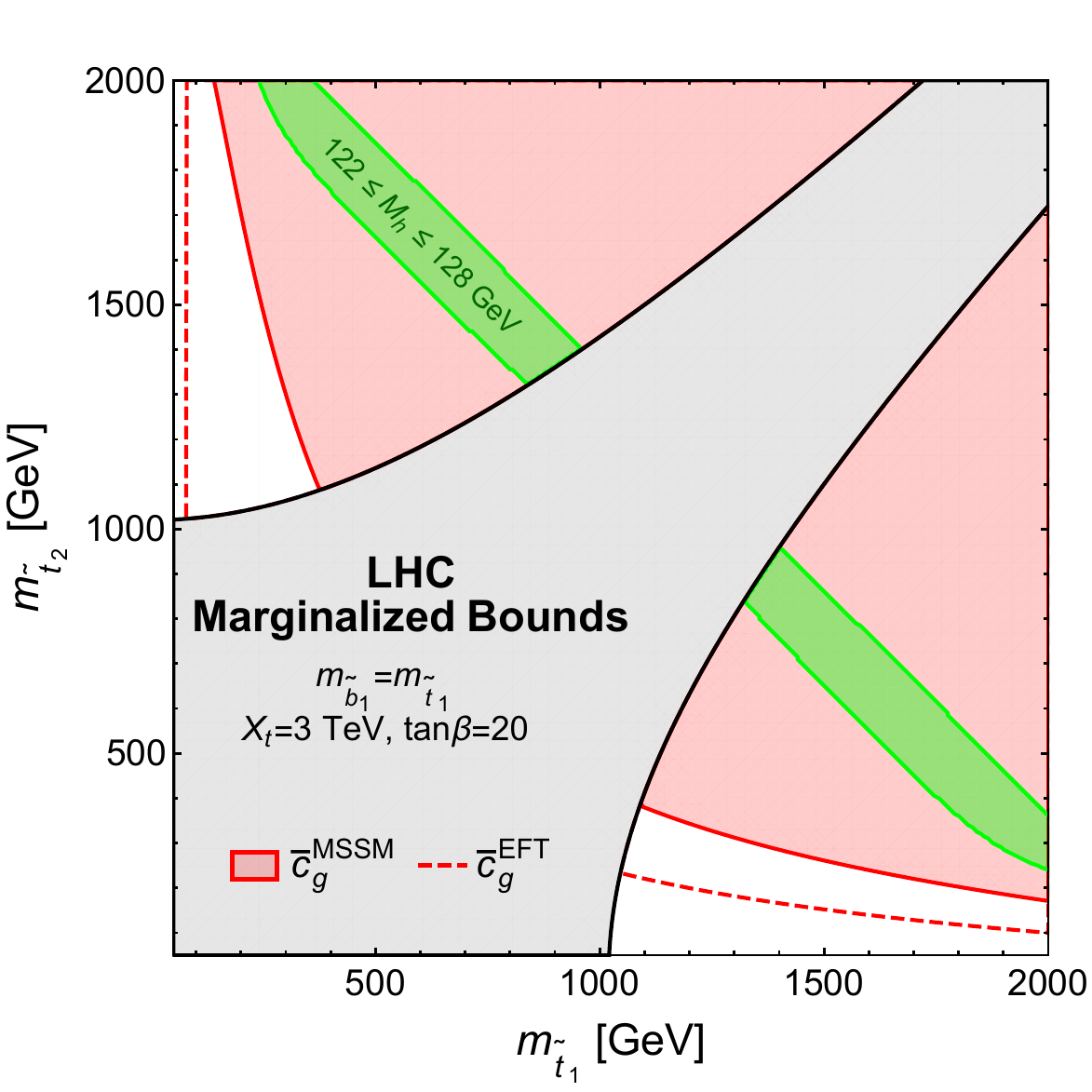}
\includegraphics[scale=0.38]{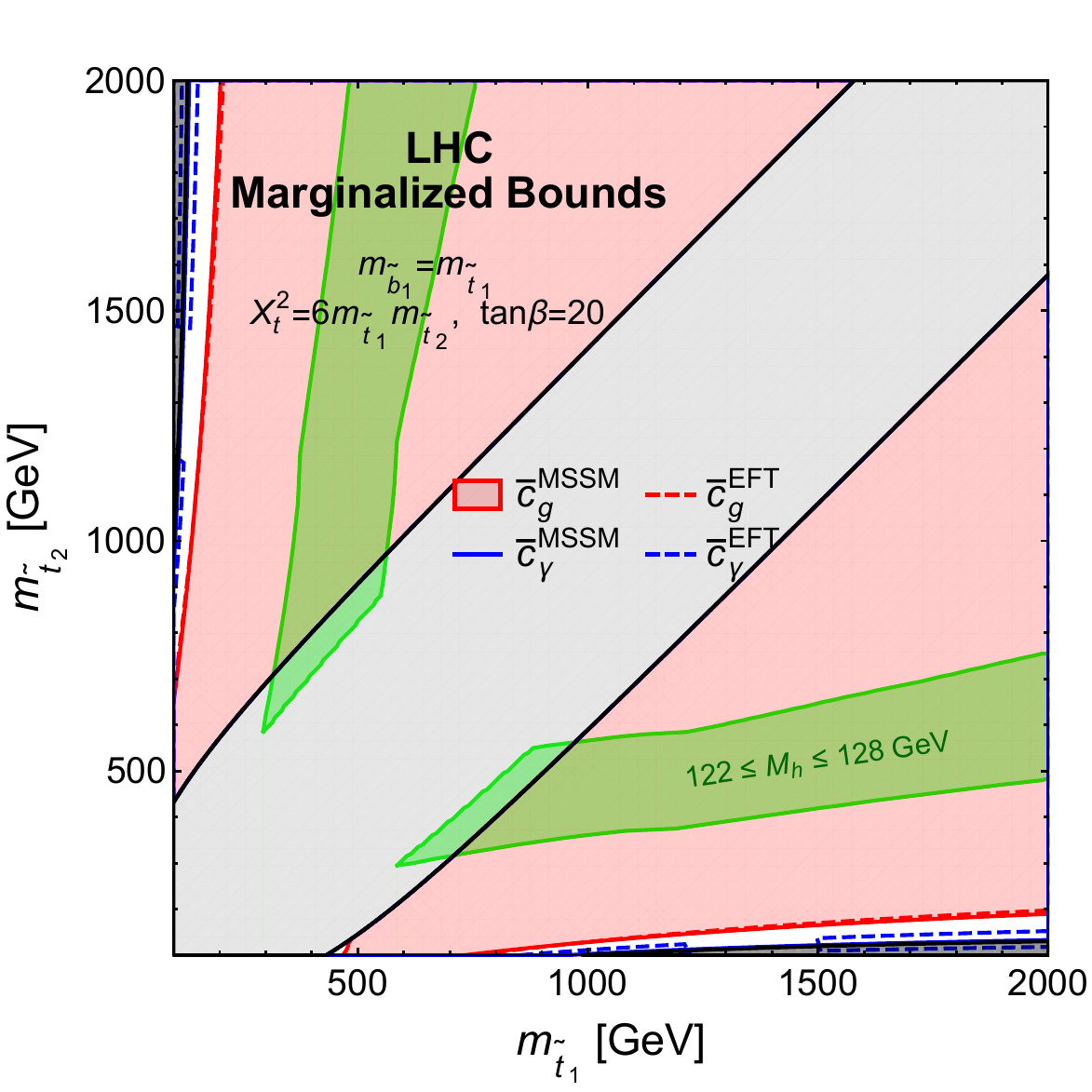}
\caption{\it Compilation of the constraints in the case of non-degenerate soft mass
parameters, including also sbottom squarks and assuming $m_{\tilde b_1} = m_{\tilde t_1}$
under the hypotheses $\tan \beta = 20$ and $X_t = 0$ (upper left panel), $X_t = 1$~TeV
(upper right panel), $X_t = 3$~TeV (lower left panel) and $X_t = \sqrt{6 m_{\tilde t_1} m_{\tilde t_2}}$
(lower right panel). The red (blue) lines show the current individual 95\% CL
constraints from $\bar{c}_g$ ($\bar{c}_\gamma$) as evaluated exactly in the MSSM (solid lines)
and in the EFT approach. Additionally, the region compatible with $\bar{c}_g$ is shaded pink,
the band compatible with $M_h$ is shaded green, and regions disallowed by the mixing
hypothesis or the appearance of a tachyonic stop are shaded grey.}
\label{fig:current_nondeg_stops}
\end{center}
\end{figure}

We see in the upper panels of Fig.~\ref{fig:current_deg_stops} that
the $\bar{c}_g$ constraints on $m_{\tilde t_1}$ are generally the strongest, 
except for large $|X_t/m_{\tilde t}|$. We also observe that the MSSM and EFT
evaluations give rather similar bounds on $m_{\tilde t_1}$ for
$|X_t/m_{\tilde t}| \lesssim 1$ and $\gtrsim 2$. However, there are significant differences
for $ 1 \lesssim |X_t/m_{\tilde t}| \lesssim 2$, due to the fact that the two evaluations
have zeroes at different values of $X_t/m_{\tilde t}$. The next most sensitive constraints
are those from $T$, parametrised here by the coefficient $\bar{c}_T$,
which become competitive with the $\bar{c}_g$ constraints at
large $|X_t/m_{\tilde t}|$, but are significantly weaker for small values of $X_t/m_{\tilde t}$.
The constraints from $\bar{c}_{\gamma}$ are weaker still for all values of $X_t/m_{\tilde t}$,
as might have been expected because the global fit in~\cite{ESY} gave constraints on $\bar{c}_{\gamma}$
that are weaker than those on $\bar{c}_g$. Indeed, the $\bar{c}_{\gamma}$ constraint is not significantly stronger
than the constraint that the ${\tilde t_1}$ not be tachyonic, as shown by the grey shading
in the upper panels of Fig.~\ref{fig:current_deg_stops}. We also note that the LHC
measurement of $M_h$ favours $|X_t/m_{\tilde t}| \gtrsim 2$ and values of $m_{\tilde t}$
that are consistent with the EFT bounds.

These results are reflected in the lower panels of Fig.~\ref{fig:current_deg_stops}, where we present
the $(m_{\tilde t_1}, m_{\tilde t_2})$ planes with the marginalized constraints (left panel) and the
individual constraints (right panel). The MSSM and EFT implementations of the $\bar{c}_g$
constraint give qualitatively similar results, and (except for extreme values of $m_{\tilde t_1}/m_{\tilde t_2}$)
are generally stronger than the constraints
from $\bar{c}_T$, which are in turn stronger than the $\bar{c}_{\gamma}$ constraint.
We also note that the LHC measurement of $M_h$ favours moderate values of
$m_{\tilde t_1}/m_{\tilde t_2}$ and values of $m_{\tilde t_1}$ or $m_{\tilde t_2} \gtrsim 520$~GeV.

The limits on the lightest stop mass for degenerate soft-supersymmetry breaking masses $m_{\tilde{Q}} = m_{\tilde{t}_R} = m_{\tilde{t}}$ with $X_t = 0$ are shown in the last column of Table~\ref{tab:currentconstraints}.

\subsection{Non-Degenerate Stop Masses}

We consider now cases with non-degenerate stop soft mass parameters, 
allowing also for the possibility that the lighter sbottom squark plays a r\^ole. We
show in Fig.~\ref{fig:current_nondeg_stops} various planes under the hypotheses
$m_{\tilde b_1} = m_{\tilde t_1}$ and $\tan \beta =20$, considering several
possibilities for $X_t$. In all panels, the constraints from the individual 95\% bound
on $\bar{c}_g$ are indicated by red lines and those from $\bar{c}_{\gamma}$ are indicated
by blue lines (solid for the exact MSSM evaluation and dashed for the EFT approach),
and the region allowed by the exact calculation is shaded pink. 

The upper left panel
is for $X_t = 0$: we see that in the limit $m_{\tilde t_2} \gg m_{\tilde t_1}$ the $\bar{c}_g$
constraint imposes $m_{\tilde t_1} \gtrsim 300$~GeV, with a difference of $\sim 20$~GeV
between the exact and EFT calculations. On the other hand, if $m_{\tilde t_2} = m_{\tilde t_1}$
we find $m_{\tilde t_1} \gtrsim 380$~GeV, again with the EFT calculation giving a bound $\sim 20$~GeV
stronger than the exact MSSM calculation. The corresponding bounds from the individual
95\% constraint on $\bar{c}_\gamma$ are $\simeq 100$~GeV weaker. However, we note that
the LHC constraint on $M_h$ is not respected anywhere in this plane.

Turning now to the case $X_t = 1$~TeV shown in the upper right panel of Fig.~\ref{fig:current_nondeg_stops},
we see a grey shaded band around the $m_{\tilde t_1} = m_{\tilde t_2}$ line that is disallowed by
${\tilde t_1} - {\tilde t_2}$ mixing, and other grey shaded regions where $m_{\tilde t_1} \ll m_{\tilde t_2}$
(or vice versa) and the lighter stop is tachyonic. In this case the $M_h$
constraint (green shaded band) can be satisfied, with small strips of the parameter space ruled out by the $\bar{c}_g$
constraint. The $\bar{c}_\gamma$ constraint is unimportant in this case.

When $X_t$ is increased to 3~TeV, as shown in the lower left panel of Fig.~\ref{fig:current_nondeg_stops},
the diagonal band forbidden by mixing expands considerably, and the $\bar{c}_\gamma$ constraint
disappears. In this case the $\bar{c}_g$ constraint would allow 
$(m_{\tilde t_1}, m_{\tilde t_2}) \gtrsim (400, 1100)$~GeV on the boundary of the band 
forbidden by the mixing hypothesis, but the $M_h$ constraint is stronger, enforcing
$(m_{\tilde t_1}, m_{\tilde t_2}) \gtrsim (800, 1300)$~GeV along this boundary.

Finally, we consider in the lower right panel of Fig.~\ref{fig:current_nondeg_stops} the so-called
maximal-mixing hypothesis $X_t = \sqrt{6m_{\tilde t_1} m_{\tilde t_2}}$. In this case,
almost the entire $(m_{\tilde t_1}, m_{\tilde t_2})$ plane is allowed by the $\bar{c}_g$
constraint, whereas a triangular region at small $m_{\tilde t_1}$ and/or $m_{\tilde t_2}$
is forbidden by the $M_h$ constraint.

It is interesting to compare the limits on $m_{\tilde t_1}$ that we find with those
found in a recent global fit to the pMSSM~\cite{MC11} in which universal third-generation
squark masses were assumed at the renormalisation scale $\sqrt{m_{\tilde t_1} m_{\tilde t_2}}$,
the first- and second-generation squark masses were assumed to be equal, but allowed to
differ from the third-generation mass as were the slepton masses,
arbitrary non-universal gaugino masses $M_{1,2,3}$ were allowed,
and the trilinear soft supersymmetry-breaking parameter $A$ was assumed to be universal
but otherwise free. That analysis included LHC, dark matter and flavour constraints, as
well as electroweak precision observables and Higgs measurements, and found 
$m_{\tilde t_1} \gtrsim 400$~GeV. The analysis of this paper uses somewhat different
assumptions and hence is not directly comparable, but it is interesting that the one-loop
sensitivity of $\bar{c}_g$ to the stop mass parameters is quite comparable.

%\newpage
\section{Sensitivities of Possible Future Precision Measurements}

\begin{table}[h!]
\begin{center}
\begin{tabular}
{|c | c | c | c  | c | c |}
\hline
\multirow{2}{*}{Coeff.} &\multicolumn{2}{|c|}{ \multirow{2}{*}{Experimental constraints}} & \multirow{2}{*}{95 \% CL limit} & \multicolumn{2}{|c|}{deg. $m_{\tilde{t}_{1}}$} \\
 & \multicolumn{2}{|c|}{} & & $X_t=0$ & $X_t=m_{\tilde{t}}/2$ \\
\hline
\multirow{4}{*}{$\bar{c}_g$} & \multirow{2}{*}{$\text{ILC}_{250 \text{GeV}}^{1150 \text{fb}^{-1}}$ } & marginalized & $[-7.7, 7.7] \times 10^{-6}$ & $\sim 675$ GeV & $\sim 520$ GeV \\ 
 & & individual  & $[-7.5, 7.5] \times 10^{-6}$ & $\sim 680$ GeV & $\sim 545$ GeV \\
 \cline{2-6}
& \multirow{2}{*}{FCC-ee} & marginalized & $[-3.0, 3.0] \times 10^{-6}$ & $\sim 1065$ GeV & $\sim 920$ GeV \\ 
 & & individual  & $[-3.0,3.0] \times 10^{-6}$ & $\sim 1065$ GeV & $\sim 915$ GeV \\
  \hline
\multirow{4}{*}{$\bar{c}_\gamma$}  & \multirow{2}{*}{$\text{ILC}_{250 \text{GeV}}^{1150 \text{fb}^{-1}}$}  & marginalized  & $[-3.4 ,3.4] \times 10^{-4}$	 & $\sim 200$ GeV & $\sim 40$ GeV \\
 & & individual  & $[-3.3, 3.3] \times 10^{-4}$ & $\sim 200$ GeV & $\sim 35$ GeV \\
\cline{2-6}
& \multirow{2}{*}{FCC-ee}  & marginalized  & $[-6.4 ,6.4] \times 10^{-5}$	 & $\sim 385$ GeV & $\sim 250$ GeV \\
 & & individual  & $[-6.3, 6.3] \times 10^{-5}$ & $\sim 390$ GeV & $\sim 260$ GeV \\
 \hline
\multirow{4}{*}{$\bar{c}_T$} & \multirow{2}{*}{$\text{ILC}_{250 \text{GeV}}^{1150 \text{fb}^{-1}}$} & marginalized & $[-3, 3] \times 10^{-4}$ & $\sim 480$ GeV & $\sim 285$ GeV \\
 & & individual  & $[-7, 7] \times 10^{-5}$ & $\sim 930$ GeV & $\sim 780$ GeV \\
\cline{2-6}
 & \multirow{2}{*}{FCC-ee} & marginalized & $[-3, 3] \times 10^{-5}$ & $\sim 1410$ GeV & $\sim 1285$ GeV \\
 & & individual  & $[-0.9, 0.9] \times 10^{-5}$ & $\sim 2555$ GeV & $\sim 2460$ GeV \\
 \hline
\multirow{4}{*}{$\bar{c}_W+\bar{c}_B$}  & \multirow{2}{*}{$\text{ILC}_{250 \text{GeV}}^{1150 \text{fb}^{-1}}$} & marginalized & $[-2, 2] \times 10^{-4}$ & $\sim 230$ GeV & $\sim 170$ GeV \\
 & & individual  & $[-6, 6] \times 10^{-5}$ & $\sim 340$ GeV & $\sim 470$ GeV \\
\cline{2-6}
& \multirow{2}{*}{FCC-ee} & marginalized & $[-2, 2] \times 10^{-5}$ & $\sim 545$ GeV & $\sim 960$ GeV \\
 & & individual  & $[-0.8, 0.8] \times 10^{-5}$ & $\sim 830$ GeV & $\sim 1590$ GeV \\
 \hline
\end{tabular}
\end{center}
\caption{\it List of the $95\%$ CL bounds on EFT operator
coefficients from projected constraints on Higgs couplings and electroweak precision observables at
the future $e^+ e^-$ colliders ILC and FCC-ee. The marginalized limits on $\bar{c}_g$ or $\bar{c}_\gamma$ 
($\bar{c}_T$ or $\bar{c}_W + \bar{c}_B$) are for a two-parameter fit allowing $\bar{c}_g$ and 
$\bar{c}_\gamma$  ($\bar{c}_T$ and $\bar{c}_W + \bar{c}_B$) to vary simultaneously but setting other operator
coefficients to zero. The corresponding lightest stop mass limits shown are for 
degenerate soft-supersymmetry breaking masses $m_{\tilde{Q}} = m_{\tilde{t}_R} = m_{\tilde{t}}$ with $X_t = 0$ and $X_t/m_{\tilde{t}}=2$.}
\label{tab:futureconstraints}
\end{table}

\begin{figure}[h!]
\begin{center} 
%\hspace*{-2.3cm}
\includegraphics[scale=0.38]{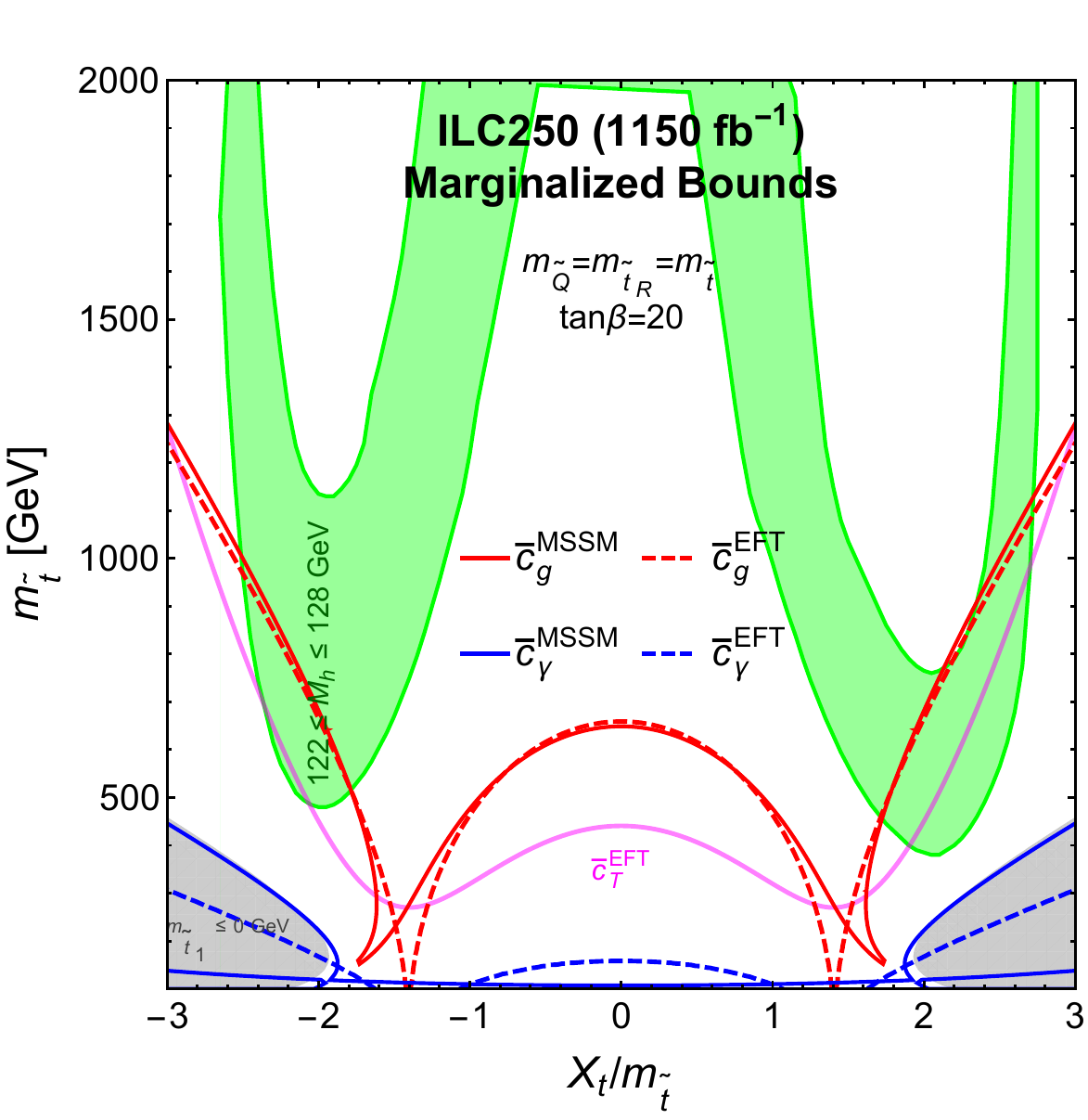}
\includegraphics[scale=0.38]{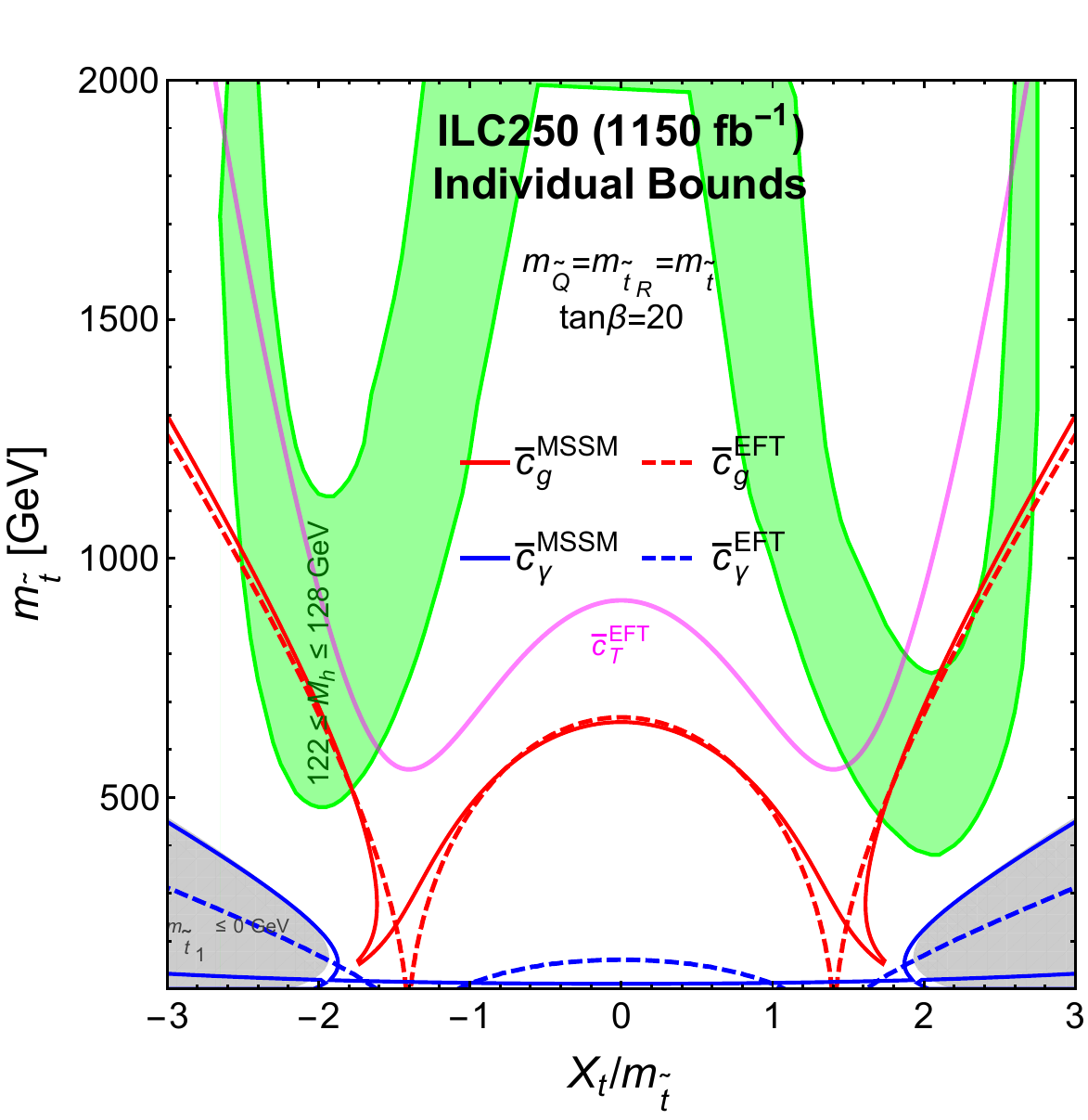} \\
\includegraphics[scale=0.38]{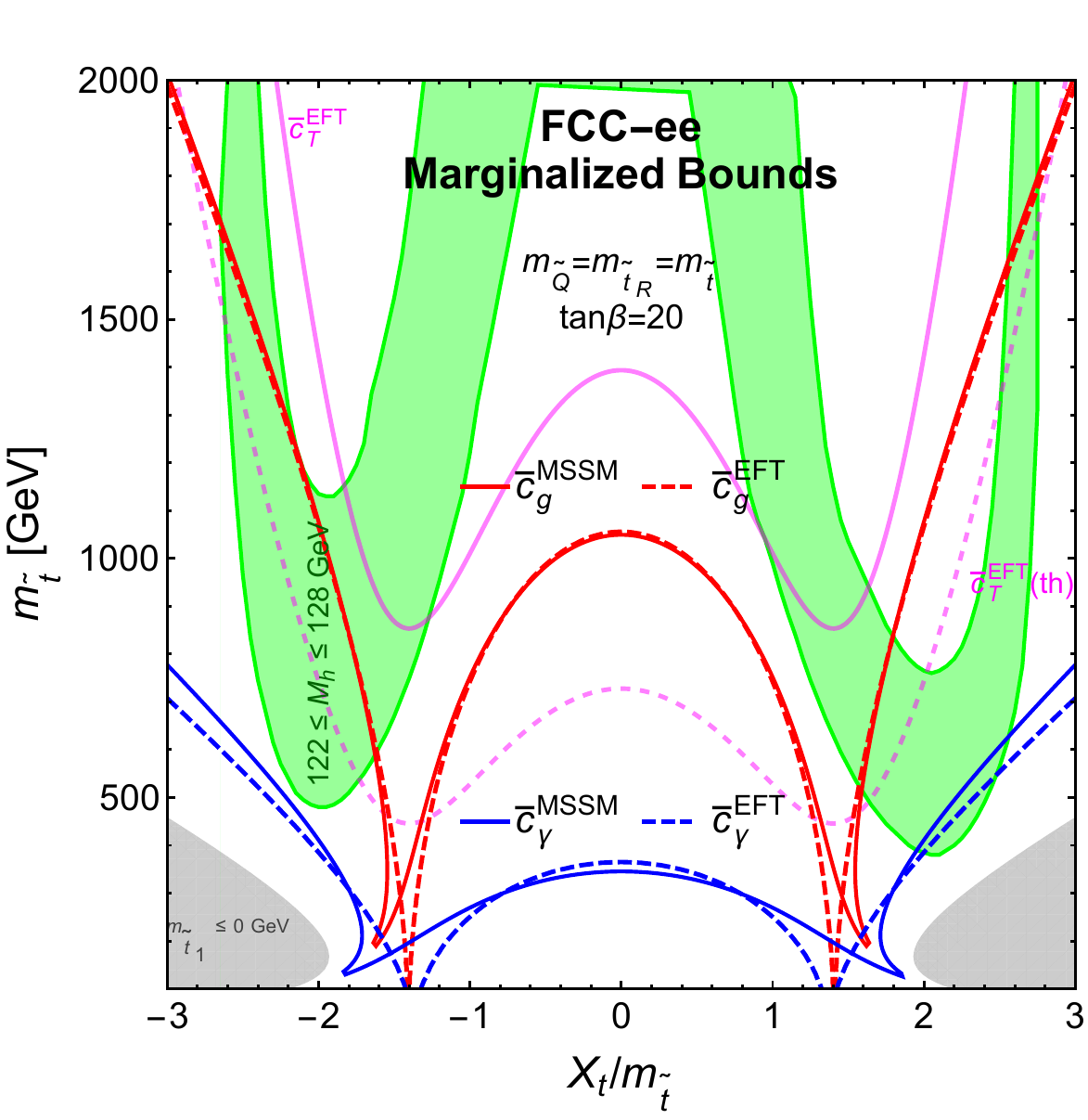}
\includegraphics[scale=0.38]{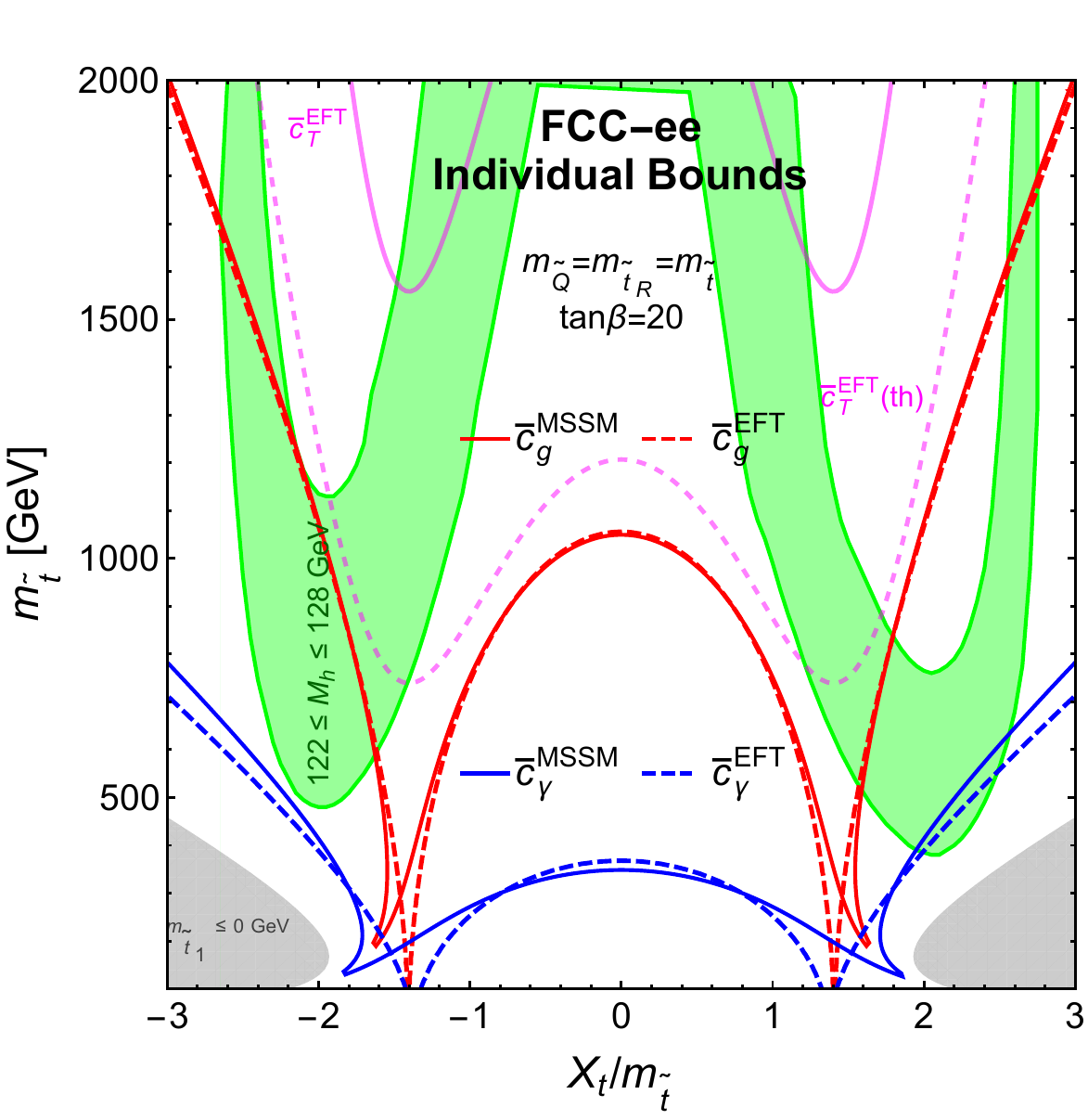}
\caption{\it The $(X_t/m_{\tilde{t}}, m_{\tilde{t}})$ planes, analogous to those in the upper panels of
Fig.~\protect\ref{fig:current_deg_stops}, showing prospective marginalized bounds (left panels) and
individual bounds (right panels) from the ILC~\protect\cite{ILCEWPT} with $1150~\text{fb}^{-1}$ of luminosity at $250$~GeV
(upper panels) and from FCC-ee~\protect\cite{blondelslides} with $10^4~\text{fb}^{-1}$ 
of luminosity at $240$~GeV (lower panels). In the latter case,
the solid purple lines are the 95\% CL contours for electroweak precision measurements
from FCC-ee incorporating the projected statistical
and systematic experimental errors alone, and the dashed purple lines also
include theory errors from~\protect\cite{mishimaslides}.
}
\label{fig:future_deg_stops}
\end{center}
\end{figure}

\begin{figure}[h!]
\begin{center} 
%\hspace*{-2.3cm}
\includegraphics[scale=0.365]{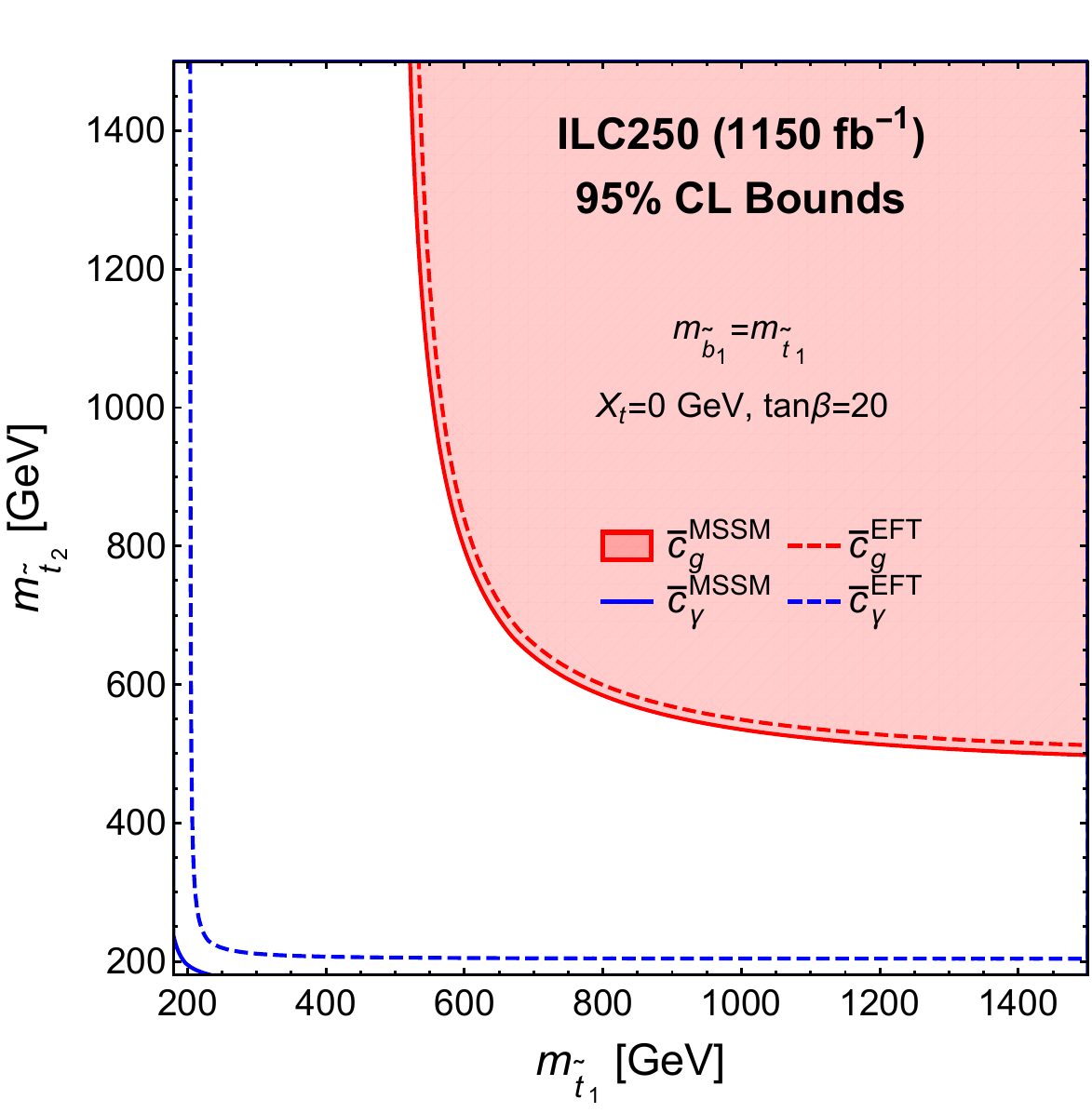}
\includegraphics[scale=0.38]{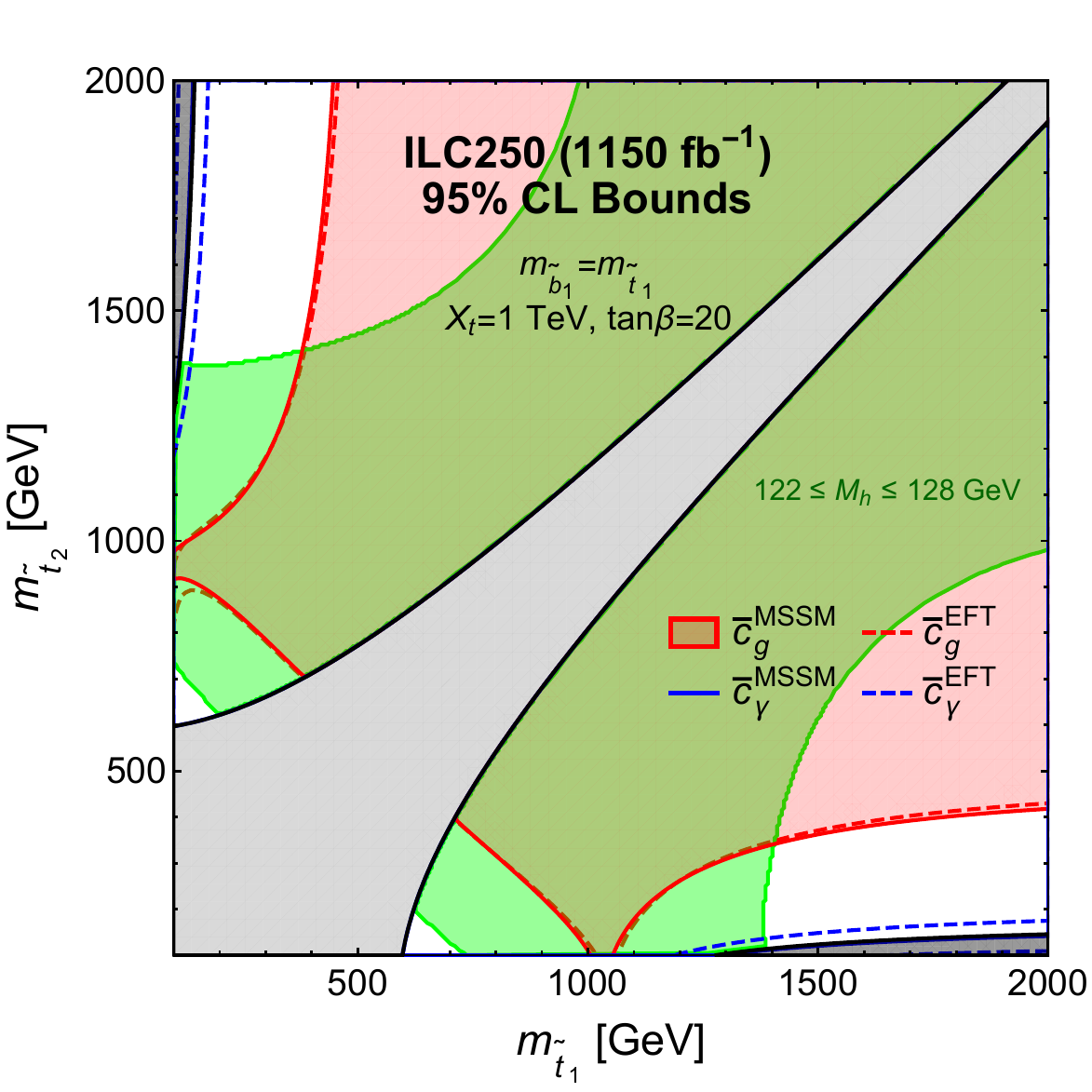}\\
\includegraphics[scale=0.38]{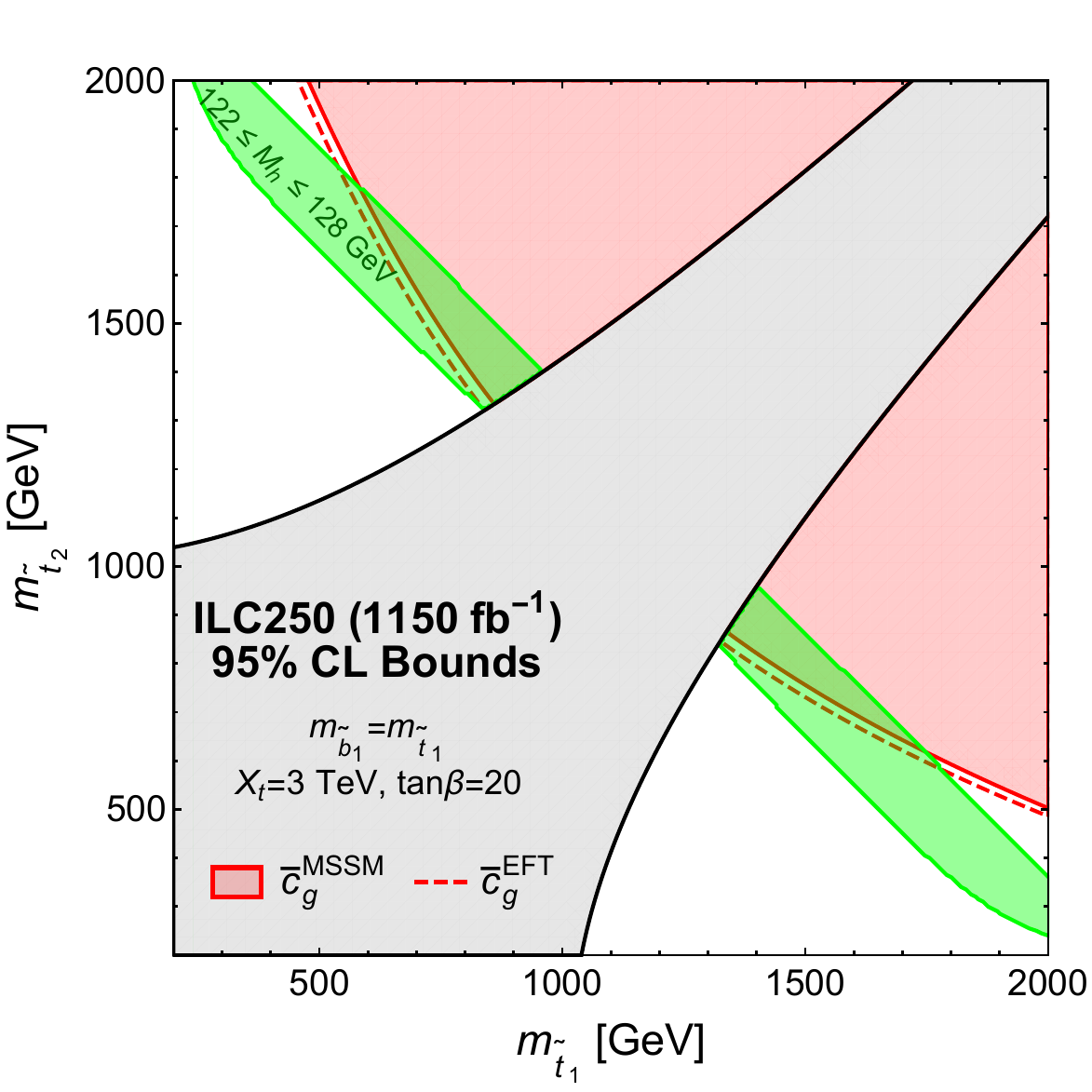}
\includegraphics[scale=0.38]{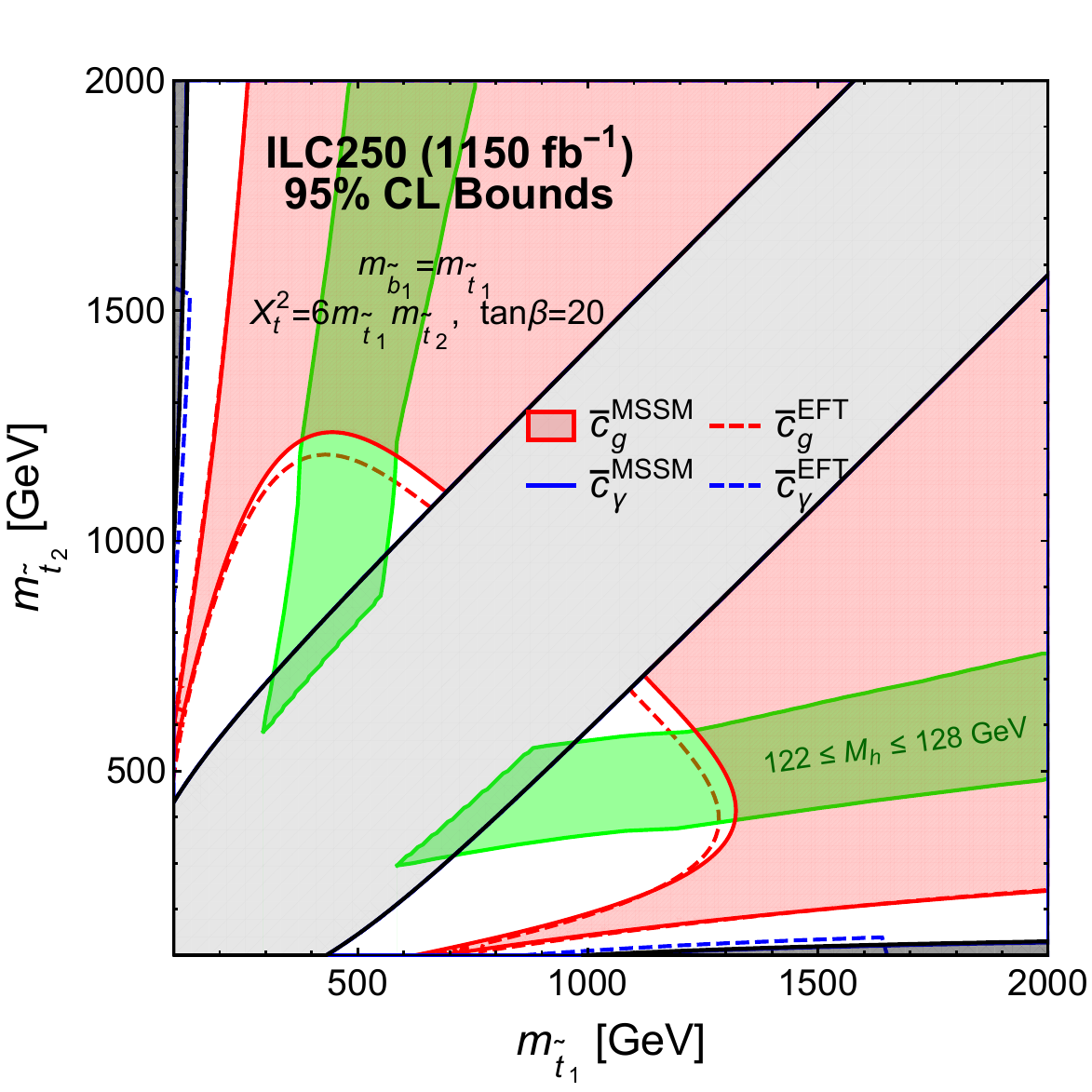}
\vspace{-0.2cm}
\caption{\it Compilation of projected ILC 95 \% CL bounds from $\bar{c}_g$ ($\bar{c}_\gamma$) given by red (blue) lines in the $m_{\tilde{t}_1}$ vs $m_{\tilde{t}_2}$ plane, analogous to Fig.~\protect\ref{fig:current_nondeg_stops}, with $m_{\tilde{b}_1}=m_{\tilde{t}_1}$ and $\tan{\beta}=20$. Values of $X_t=0,1,3,\sqrt{6m_{\tilde{t}_1}m_{\tilde{t}_2}}$ TeV are shown clockwise from top left. The marginalized limits are displayed and the individual bounds are very similar.}
\label{fig:ILC_nondeg_stops}
\end{center}
\end{figure}

\begin{figure}[h!]
\begin{center} 
%\hspace*{-2.3cm}
\includegraphics[scale=0.378]{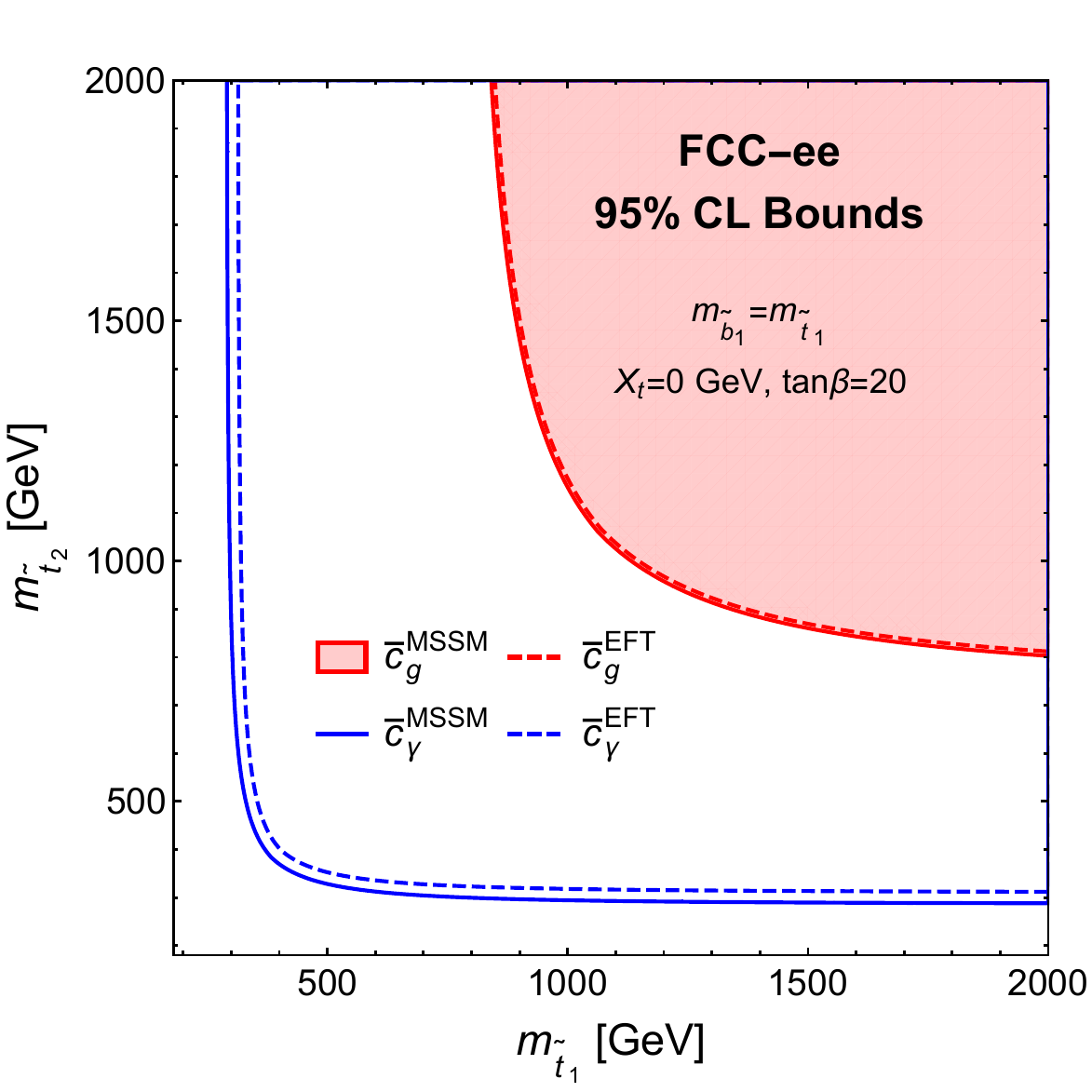}
\includegraphics[scale=0.38]{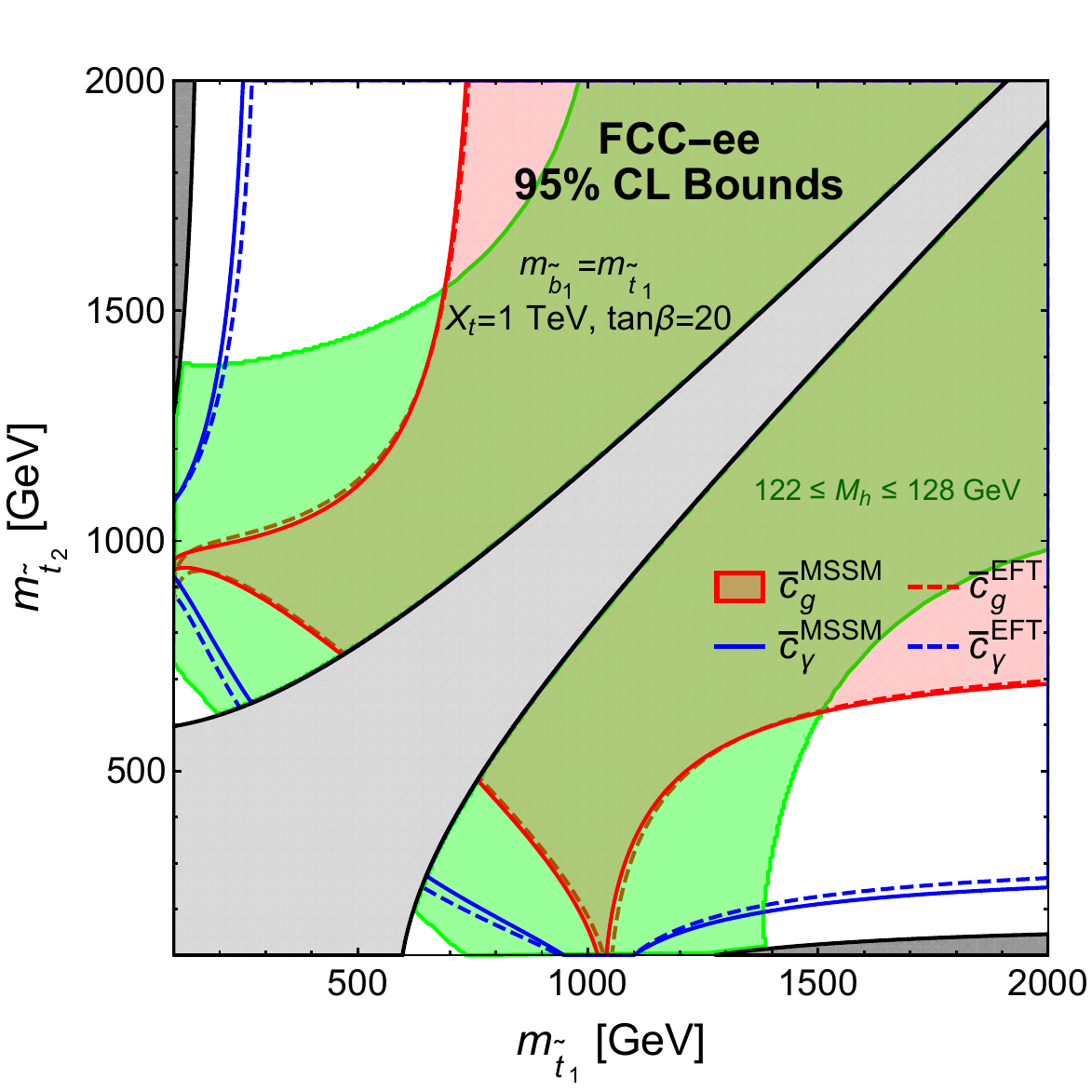}\\
\includegraphics[scale=0.38]{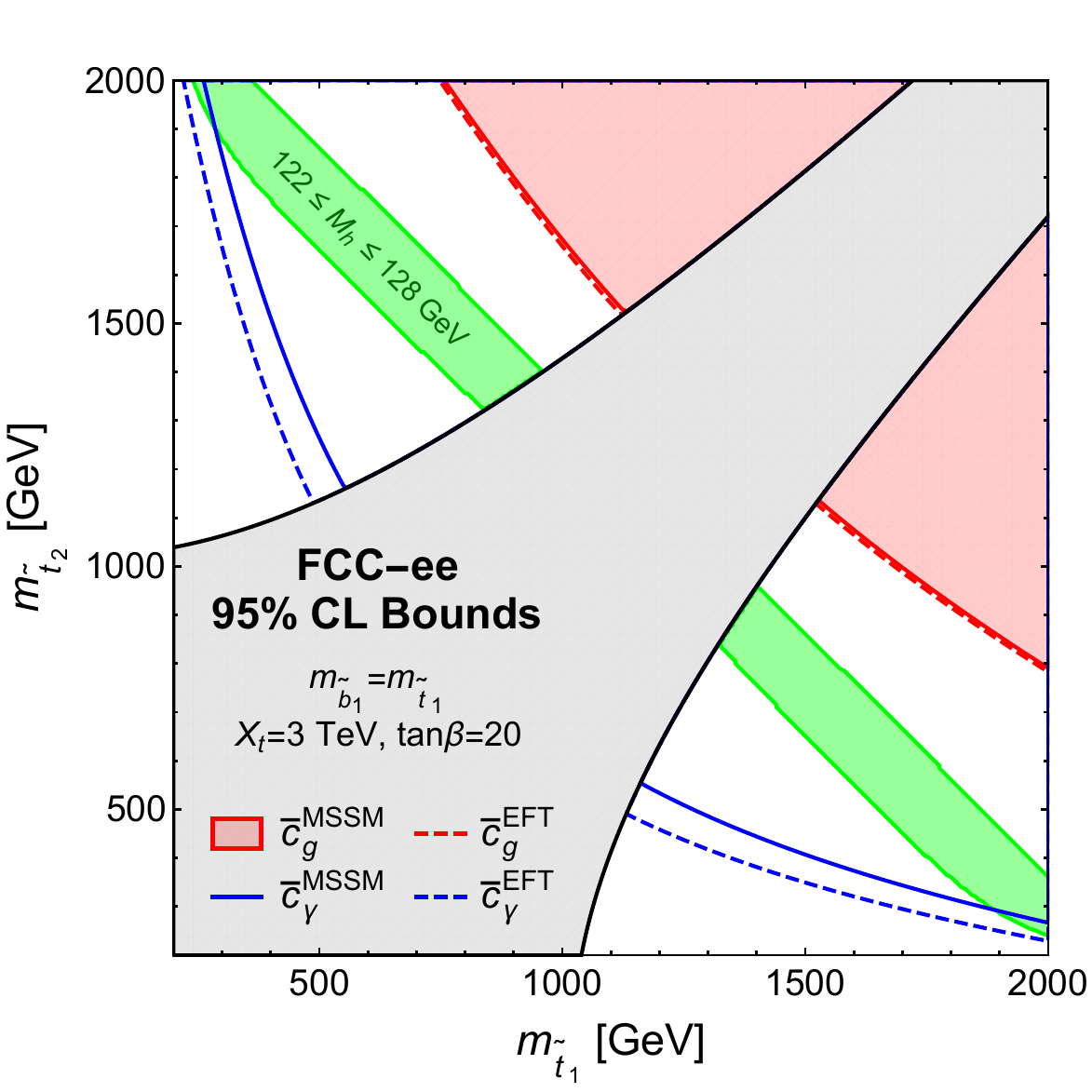}
\includegraphics[scale=0.38]{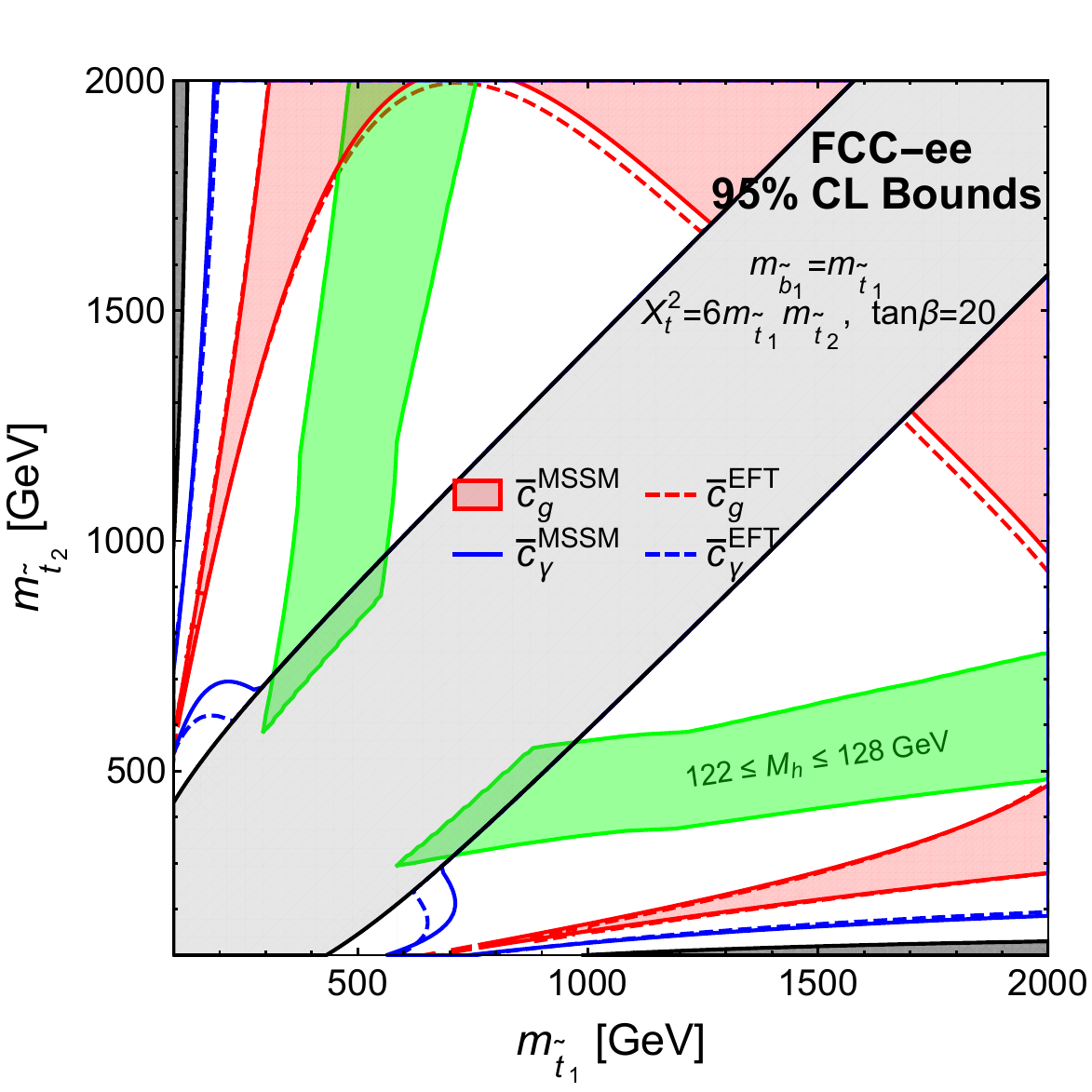}
\vspace{-0.2cm}
\caption{\it Compilation of projected FCC-ee 95 \% CL bounds from $\bar{c}_g$ ($\bar{c}_\gamma$) given by red (blue) lines in the $m_{\tilde{t}_1}$ vs $m_{\tilde{t}_2}$ plane, analogous to Fig.~\protect\ref{fig:current_nondeg_stops}, with $m_{\tilde{b}_1}=m_{\tilde{t}_1}$ and $\tan{\beta}=20$. Values of $X_t=0,1,3,\sqrt{6m_{\tilde{t}_1}m_{\tilde{t}_2}}$ TeV is shown clockwise from top left. The marginalized limits are displayed and the individual bounds are very similar.  }
\label{fig:FCCee_nondeg_stops}
\end{center}
\end{figure}

We saw in the previous Section that the precision of current measurements does not exclude in a 
model-independent way most of the parameter space for a stop below the TeV scale, 
and barely reaches into the region required for a 125 GeV Higgs mass in the MSSM. 
However, future colliders will increase significantly the precision of electroweak and Higgs measurements
to the level required to challenge seriously the naturalness paradigm and test the MSSM calculations
of $M_h$.

In this Section we assess the 
potential improvements for constraints on a light stop possible with future $e^+ e^-$
colliders. As previously, we perform an analysis in the 
EFT framework via the corresponding bounds on the relevant dimension-6 coefficients,
and compare it with the exact one-loop MSSM calculation. 
As representative examples of future $e^+ e^-$
colliders, we focus on the ILC~\cite{ILCEWPT} and FCC-ee~\cite{blondelslides} (formerly known as TLEP) proposals. 
The scenarios considered here for the ILC and FCC-ee postulate centre-of-mass energies of
250 and 240~GeV with luminosities of 1150 $\text{fb}^{-1}$ and 10000 $\text{fb}^{-1}$, respectively. 

Table~\ref{tab:futureconstraints} lists the prospective 95\% CL limits obtained on $\bar{c}_g, \bar{c}_\gamma, \bar{c}_T,$ 
and $\bar{c}_W+\bar{c}_B$ from a $\chi^2$ analysis, with the marginalized constraints on 
$\bar{c}_g$ and $\bar{c}_\gamma$ obtained in a two-parameter fit to just these coefficients, 
and similarly for $\bar{c}_T$ and $\bar{c}_W+\bar{c}_B$, corresponding to the $T$ and $S$ 
parameters respectively, as well as the constraints obtained when each operator coefficient is
allowed individually to be non-zero. The target precisions on experimental errors for the 
electroweak precision observables $m_W, \Gamma_Z, R_l$ and $A_l$ at the ILC are given in \cite{ILCEWPT}, 
and those at FCC-ee were taken from~\cite{blondelslides}, and include important systematic uncertainties. 
The errors on the Higgs associated production cross-section times branching ratio are
from~\cite{ILCwhitepaper} for the ILC and from~\cite{FCC-ee} for FCC-ee. 
The numbers quoted in Table~\ref{tab:futureconstraints} neglect theoretical uncertainties,
in order to reflect the possible performances of the experiments~\footnote{We also show
as dashed purple lines in the FCC-ee panels the weaker constraints obtained using the estimates
of theoretical uncertainties in~\cite{mishimaslides}, while noting that these have not been studied in detail.}.
The treatment of the dimension-6 coefficients in the observables follows a
procedure similar to that of the global fit performed in~\cite{ESY}, and we use the results of~\cite{craigetal}
to rescale the constraint from associated Higgs production.

\subsection{Degenerate Stop Masses}

Contours from possible future constraints on $\bar{c}_g$, $\bar{c}_\gamma$ and $\bar{c}_T$ 
for the case of degenerate soft masses 
$m_{\tilde{Q}}=m_{\tilde{t}_R}\equiv m_{\tilde{t}}$ are plotted in Fig.~\ref{fig:future_deg_stops},
using again the value $\tan{\beta}=20$. 
The upper panels show results for the ILC, the lower panels for FCC-ee, the left panels show the marginalized
constraints and the right panels show the individual constraints.
The grey and green shaded regions are the same as in Fig.~\ref{fig:future_deg_stops}.
We see that the marginal and individual sensitivities to $m_{\tilde t}$ from $\bar{c}_g$ and $\bar{c}_\gamma$
are very similar, whereas the individual sensitivity of $\bar{c}_T$ are much stronger, particularly at FCC-ee. 
We see that ILC is indirectly sensitive to $m_{\tilde t} \sim 600$~GeV, 
and that FCC-ee is indirectly sensitive to stops in the TeV range.
The measurement of the $\bar{c}_T$ coefficient at FCC-ee has the highest potential reach, 
though this will be highly dependent on future improvements in reducing theory uncertainties~\cite{FCC-ee, mishimaslides}. 

The limits on the lightest stop mass for degenerate soft-supersymmetry breaking masses $m_{\tilde{Q}} = m_{\tilde{t}_R} = m_{\tilde{t}}$ with $X_t = 0$ and $X_t/ m_{\tilde{t}}=2$ are shown in the two last columns of Table~\ref{tab:futureconstraints}.

\subsection{Non-Degenerate Stop Masses}

Moving on to the non-degenerate case, the $\bar{c}_g$ and $\bar{c}_\gamma$ 95\% CL limits for 
ILC and FCC-ee are plotted in the $m_{\tilde{t}_1}$ vs $m_{\tilde{t}_2}$ plane for various  $X_t$ 
values in Fig.~\ref{fig:ILC_nondeg_stops} and \ref{fig:FCCee_nondeg_stops} respectively. 
The top left, top right, and bottom left plots correspond to $X_t = 0, 1$ and 3 TeV respectively, 
while the bottom right plot is for the maximal-mixing hypothesis $X_t = \sqrt{6 m_{\tilde{t}_1}m_{\tilde{t}_2}}$. 
We see that the ILC sensitivity to $\bar{c}_g$ begins to probe and potentially
exclude parts of the green shaded region compatible with the measured $M_h$, 
while FCC-ee would push the sensitivity of $\bar{c}_g$ constraints into the TeV scale. 
In particular, it could eliminate the entire allowed $M_h$ region for $X_t=3$ TeV.

\section{Conclusions and Prospects}

In light of the SM-like Higgs sector and the current lack of direct evidence for additional degrees of freedom beyond the SM,
the framework of the Effective SM (ESM) is gaining increasing attention as a general framework for
 characterising the indirect effects of possible new physics in a model-independent way. 
 The ESM is simply the SM extended in the way it has always been regarded: 
 as an effective field theory supplemented by higher-dimensional operators suppressed by the scale of new physics. 
 The leading lepton-number-conserving effects are parametrised by dimension-6 operators, 
 whose coefficients are determined by matching to a UV model and constrained through their effects on 
 experimental observables. In this paper we have illustrated all these steps in the EFT approach for light stops in the MSSM. 

In particular, we employed the CDE method to compute the one-loop effective Lagrangian, 
showing how certain results derived previously under the assumption of a degenerate 
mass matrix can be generalised to the non-degenerate case. The universal one-loop 
effective Lagrangian can then be used without caveats to obtain directly one-loop Wilson coefficients. 
The advantage of this was demonstrated here in the calculation of the $\bar{c}_g$ and 
$\bar{c}_\gamma$ coefficients. One simply takes the mass and $U$ matrices from the 
quadratic term of the heavy field being integrated out, as defined in (\ref{eq:lagrangianUV}),
and substitutes it with the corresponding field strength matrix into the universal expression obtained from the CDE expansion~\cite{DEQY2} to get the desired operators,
 without having to evaluate any loop integrals or match separate calculations in the UV and EFT.

Since the $hgg$ and $h\gamma\gamma$ couplings are loop-induced in the SM, 
the $\bar{c}_g$ and $\bar{c}_\gamma$ coefficients are currently the most sensitive to light stops. 
The stop contribution to these coefficients is also loop-suppressed, thus lowering the EFT cut-off scale, and it is natural to ask at what point the EFT breaks down and the effects of higher-dimensional operators 
are no longer negligible. We addressed this question by comparing the EFT coefficients with a full calculation in the 
MSSM, finding that the disagreement is generally $\lesssim 10 \%$ for a lightest stop mass $m_{\tilde{t}_1} \gtrsim 500$ GeV, 
with the exception of a large $|X_t| \geq 3 m_{\tilde{t}_1}$ or accidental cancellations in the Higgs-stop couplings. 

The constraints on $\bar{c}_g$ and $\bar{c}_\gamma$ from a global fit to the current
LHC and Tevatron data, and the constraints on $\bar{c}_T$ and $\bar{c}_W+\bar{c}_B$ from
LEP electroweak precision observables, were then translated into the corresponding constraints on the 
stop masses and $X_t$. The coefficient $\bar{c}_g$ is the most sensitive, followed by $\bar{c}_T$,
which is equivalent to the oblique $T$ parameter. In the case of degenerate soft masses, this analysis requires $m_{\tilde{t}_1} \gtrsim 410$ GeV for $X_t=0$, and $m_{\tilde{t}_1} \gtrsim 200$ GeV if we also apply the Higgs mass constraint. This is competitive with direct searches and is complimentary in the sense that it does not depend on 
how the stop decays. The limits in the non-degenerate case are generally weaker than the 
Higgs mass requirement, though a few strips in the parameter space compatible with $M_H$ can still be excluded. 

The sensitivity of future colliders can greatly improve the reach of indirect constraints into the region of 
parameter space compatible with the observed Higgs mass. The most promising measurements will be the 
$hgg$ coupling and the $T$ parameter, with FCC-ee capable of reaching a sensitivity to stop masses above 1 TeV.
Thus, FCC-ee measurements will be able to challenge the naturalness paradigm in a rather
model-independent way.

As LHC Run 2 gets under way, the question how to interpret any new physics or lack thereof
will be aided by the systematic approach of the ESM. We have demonstrated this for the case of 
light stops in the MSSM, showing how the EFT framework can simplify both the calculation of 
relevant observables and the application of experimental constraints on these observables,
giving results similar to exact one-loop calculations in the MSSM. 

%\newpage

\section*{Acknowledgements}

The work of AD was supported by the STFC Grant ST/J002798/1. 
The work of JE was supported partly by the London Centre
for Terauniverse Studies (LCTS), using funding from the European Research Council via the Advanced Investigator
Grant 26732, and partly by the STFC Grants ST/J002798/1 and ST/L000326/1.
The work of JQ was supported by the STFC Grant ST/L000326/1. 
The work of TY was supported by a
Graduate Teaching Assistantship from King's College London.

 \providecommand{\href}[2]{#2}\begingroup\raggedright

\end{document}